\documentclass[twocolumn,showpacs,preprintnumbers,amsmath,amssymb,superscriptaddress]{revtex4}

\usepackage{graphicx}
\usepackage{dcolumn}
\usepackage{bm}

\def\bea{\begin{eqnarray}}
\def\eea{\end{eqnarray}}

\begin{document}



\preprint{Version 4.0}

\title{The multiplicity dependence of inclusive $p_t$ spectra \\ from p-p collisions at $\sqrt{s}$ = 200 GeV}
\affiliation{}
\affiliation{Argonne National Laboratory, Argonne, Illinois 60439}
\affiliation{University of Birmingham, Birmingham, United Kingdom}
\affiliation{Brookhaven National Laboratory, Upton, New York 11973}
\affiliation{California Institute of Technology, Pasadena, California 91125}
\affiliation{University of California, Berkeley, California 94720}
\affiliation{University of California, Davis, California 95616}
\affiliation{University of California, Los Angeles, California 90095}
\affiliation{Carnegie Mellon University, Pittsburgh, Pennsylvania 15213}
\affiliation{Creighton University, Omaha, Nebraska 68178}
\affiliation{Nuclear Physics Institute AS CR, 250 68 \v{R}e\v{z}/Prague, Czech Republic}
\affiliation{Laboratory for High Energy (JINR), Dubna, Russia}
\affiliation{Particle Physics Laboratory (JINR), Dubna, Russia}
\affiliation{University of Frankfurt, Frankfurt, Germany}
\affiliation{Institute of Physics, Bhubaneswar 751005, India}
\affiliation{Indian Institute of Technology, Mumbai, India}
\affiliation{Indiana University, Bloomington, Indiana 47408}
\affiliation{Institut de Recherches Subatomiques, Strasbourg, France}
\affiliation{University of Jammu, Jammu 180001, India}
\affiliation{Kent State University, Kent, Ohio 44242}
\affiliation{Institute of Modern Physics, Lanzhou, China}
\affiliation{Lawrence Berkeley National Laboratory, Berkeley, California 94720}
\affiliation{Massachusetts Institute of Technology, Cambridge, MA 02139-4307}
\affiliation{Max-Planck-Institut f\"ur Physik, Munich, Germany}
\affiliation{Michigan State University, East Lansing, Michigan 48824}
\affiliation{Moscow Engineering Physics Institute, Moscow Russia}
\affiliation{City College of New York, New York City, New York 10031}
\affiliation{NIKHEF and Utrecht University, Amsterdam, The Netherlands}
\affiliation{Ohio State University, Columbus, Ohio 43210}
\affiliation{Panjab University, Chandigarh 160014, India}
\affiliation{Pennsylvania State University, University Park, Pennsylvania 16802}
\affiliation{Institute of High Energy Physics, Protvino, Russia}
\affiliation{Purdue University, West Lafayette, Indiana 47907}
\affiliation{Pusan National University, Pusan, Republic of Korea}
\affiliation{University of Rajasthan, Jaipur 302004, India}
\affiliation{Rice University, Houston, Texas 77251}
\affiliation{Universidade de Sao Paulo, Sao Paulo, Brazil}
\affiliation{University of Science \& Technology of China, Hefei 230026, China}
\affiliation{Shanghai Institute of Applied Physics, Shanghai 201800, China}
\affiliation{SUBATECH, Nantes, France}
\affiliation{Texas A\&M University, College Station, Texas 77843}
\affiliation{University of Texas, Austin, Texas 78712}
\affiliation{Tsinghua University, Beijing 100084, China}
\affiliation{Valparaiso University, Valparaiso, Indiana 46383}
\affiliation{Variable Energy Cyclotron Centre, Kolkata 700064, India}
\affiliation{Warsaw University of Technology, Warsaw, Poland}
\affiliation{University of Washington, Seattle, Washington 98195}
\affiliation{Wayne State University, Detroit, Michigan 48201}
\affiliation{Institute of Particle Physics, CCNU (HZNU), Wuhan 430079, China}
\affiliation{Yale University, New Haven, Connecticut 06520}
\affiliation{University of Zagreb, Zagreb, HR-10002, Croatia}

\author{J.~Adams}\affiliation{University of Birmingham, Birmingham, United Kingdom}
\author{M.M.~Aggarwal}\affiliation{Panjab University, Chandigarh 160014, India}
\author{Z.~Ahammed}\affiliation{Variable Energy Cyclotron Centre, Kolkata 700064, India}
\author{J.~Amonett}\affiliation{Kent State University, Kent, Ohio 44242}
\author{B.D.~Anderson}\affiliation{Kent State University, Kent, Ohio 44242}
\author{M.~Anderson}\affiliation{University of California, Davis, California 95616}
\author{D.~Arkhipkin}\affiliation{Particle Physics Laboratory (JINR), Dubna, Russia}
\author{G.S.~Averichev}\affiliation{Laboratory for High Energy (JINR), Dubna, Russia}
\author{Y.~Bai}\affiliation{NIKHEF and Utrecht University, Amsterdam, The Netherlands}
\author{J.~Balewski}\affiliation{Indiana University, Bloomington, Indiana 47408}
% ...

\collaboration{STAR Collaboration}

\date{\today}

\begin{abstract}
We report measurements of transverse momentum $p_t$ spectra for ten event multiplicity classes of p-p collisions at $\sqrt{s} = 200$~GeV. By analyzing the multiplicity dependence we find that the spectrum shape can be decomposed into a part with amplitude proportional to multiplicity and described by a L\'evy distribution on transverse mass $m_t$, and a part with amplitude proportional to multiplicity squared and described by a gaussian distribution on transverse rapidity $y_t$. The functional forms of the two parts are nearly independent of event multiplicity. The two parts can be identified with the soft and hard components of a two-component model of p-p collisions. This analysis then provides the first isolation of the hard component of the $p_t$ spectrum as a distribution of simple form on $y_t$.
\end{abstract}

\pacs{24.40.Ep,24.60-k,24.85+p,25.40.Ve,25.75.Gz}
\keywords{$p_t$ spectra, p-p collisions, two-component model, power-law model}

\maketitle

\section{Introduction}


The structure of the inclusive $p_t$ spectrum from relativistic nuclear collisions is affected by several aspects of collision dynamics and by the final-state hadronization process. Comparisons of p-p, d-Au and Au-Au $p_t$ spectra at RHIC suggest that a form of color-deconfined matter has been created in Au-Au collisions~\cite{pptrack,specphys}. Particle production mechanisms which could determine spectrum structure include soft parton scattering followed by longitudinal or `string' fragmentation~\cite{lund} and hard parton scattering followed by transverse fragmentation~\cite{hardscatt}. Other mechanisms could be significant.  The structure of the $p_t$ spectrum at some achievable level of precision may therefore be complex. A summary of efforts to unfold and interpret the structure of inclusive $p_t$ spectra from ISR to Fermilab and SP\=PS energies in the context of jet phenomenology and QCD (quantum chromodynamic) theory is provided in~\cite{mjtann}. 

At RHIC energies hard parton scattering is expected to dominate the spectrum at larger $p_t$ and to be significantly modified in A-A collisions (jet quenching)~\cite{hardscatt,transfrag}. But how does hard scattering contribute at smaller $p_t$? How does it interact with thermal or `soft'  particle production? Is there an `intermediate' $p_t$ region~\cite{interpt} with its own unique production mechanisms? Those issues remain unresolved after much theoretical speculation and experimental measurement and provide a context for the present analysis applied to high-statistics $p_t$ spectra from ten multiplicity classes of p-p collisions. The multiplicity dependence offers new access to underlying particle production mechanisms.


$p_t$ spectra from relativistic nuclear collisions are conventionally modeled by the {\em power-law} function~\cite{ua1}, a form suggested by measured jet systematics and perturbative QCD (pQCD) expectations. At larger $p_t$ the spectrum is expected to tend asymptotically to the power-law form $p_t^{-n}$~\cite{pqcd}. The strict power-law form  is then generalized to the function $A/(1+p_t / p_0)^n$, having the expected pQCD dependence at larger $p_t$ but transitioning to an approximate Maxwell-Boltzmann form at smaller $p_t$, consistent with expectations for thermal particle production. Although the power-law function has been previously applied to p-p data with apparently good fit quality ($\chi^2$ within expected limits) it has not been tested with the precision of recently-acquired RHIC p-p data. One can question the validity of its underlying assumptions. For instance, why should a single model function adequately describe spectra which may represent a mixture of several particle production mechanisms?


Alternatively, a model function can be formulated in terms of the {\em two-component} model of nuclear collisions~\cite{tcomp1}, which identifies `soft' p-p collisions with no hard parton scatter and `semi-hard' collisions with at least one significant parton scatter ({\em i.e.,} producing distinguishable hadron fragments).  According to the two-component model the minimum-bias distribution on event multiplicity $n_{ch}$ can be decomposed into separate negative binomial distributions (NBD) identified with soft and semi-hard event types. We then expect the {\em fraction} of events with a hard parton collision to increase monotonically with selected event multiplicity $n_{ch}$. Variation of $p_t$ spectra with $n_{ch}$ could then provide a basis for isolating soft and hard (and possibly other) components of inclusive  spectra on a statistical basis, where the hard spectrum component refers to the { fragment} $p_t$ spectrum for hard-scattered partons, and the soft component is the $p_t$ spectrum for `soft' particle production. 


In this analysis we first test the ability of the conventional power-law model function to represent the data. We then reconsider the data with no {\em a priori} assumptions. We attempt to describe all spectrum structure with the simplest algebraic model required by the data ({\em e.g.,} `simple' in terms of parameter number and functional forms - {\em cf.} Eq.~(\ref{2compalg}) and Sec.~\ref{discussion}) and then to associate the model elements with possible particle production mechanisms. We adopt two new analysis techniques: 1) We introduce transverse rapidity $y_t$~\cite{yt,carr} as an alternative to $p_t$. $y_t$ has the advantage that spectrum structure associated with hard parton scattering and fragmentation is {\em more uniformly represented} on a logarithmic variable: $y_t$ corresponds to variable $\xi_p = \ln (p_{\text{parton}} / p_{\text{fragment}})$ conventionally used to describe parton fragmentation functions in elementary collisions~\cite{fragfunc}. A simple description of soft particle production is {not} compromised by the choice of transverse rapidity. 2) We introduce the running integral of the $y_t$ spectrum, which substantially reduces statistical fluctuations relative to significant structure and therefore improves the precision of the analysis. 


In this paper we present high-statistics $p_t$ spectra for ten multiplicity classes from  p-p collisions at $\sqrt{s} = 200$ GeV. We use the conventional power-law model function to fit those spectra and assess the quality of that description. We then construct running integrals of the spectra on $y_t$ and define a reference function common to all $n_{ch}$ values and based on the L\'evy distribution. We use that reference to extract difference spectra which contain the $n_{ch}$-dependent parts of the spectra in a more differential form.  We find that the difference spectra have a simple structure: the major component is well-described by a gaussian distribution with fixed shape and with amplitude (relative to the reference) linearly proportional to the particle multiplicity. To simplify presentation we initially describe approximate relationships and optimized parameters without errors. We then return to a comprehensive discussion of the parameter system and its errors and consistency in Sec.~\ref{twocomp1}. This analysis is based on p-p collisions at $\sqrt{s} = 200$~GeV observed with the STAR detector at the Relativistic Heavy Ion Collider (RHIC).

\section{$p_t$ and $y_t$ Spectra} 

Data for this analysis in the form of inclusive $p_t$ spectra for unidentified charged particles were obtained from non-single-diffractive (NSD) p-p collisions at $\sqrt{s} = 200$~GeV triggered by a coincidence of two beam-beam counters (BBC) in $3.3 < |\eta| < 5$~\cite{pptrack}. Charged particles were measured with the STAR Time Projection Chamber (TPC) and Central Trigger Barrel (CTB)~\cite{starnim}. Particle momenta were determined with a 0.5 T magnetic field parallel to the beam (z) axis.  Primary charged particles were represented by TPC tracks falling within the acceptance for this analysis -- $2\pi$ azimuth, pseudorapidity $|\eta| < 0.5$, and $0.2 < p_t < 6$~GeV/c -- and satisfying track cuts described in~\cite{pptrack}. The {\em observed} particle multiplicity in the acceptance is denoted by $\hat n_{ch}$, whereas the corrected and $p_t$-extrapolated {\em true} event multiplicity is denoted by $n_{ch}$. From 3 x 10$^6$ NSD events individual $p_t$ distributions were formed for 10 primary-particle multiplicity classes indexed by the {observed} multiplicity: $\hat n_{ch} \in [1,\cdots,8,9+10,11+12]$.



To eliminate backgrounds from event pileup 
each TPC primary-track candidate was independently required to match a CTB/trigger timing requirement (100 ns, matching efficiency 94\%, false-coincidence background 2\%) and project to the beam line within 1 cm transverse distance of closest approach. No other vertex requirement was applied to the primary tracks. The event-vertex $z$ position was estimated by the arithmetic mean $\bar z$ of projected track $z$ for all CTB-matched primary tracks in an event. Events with $|\bar z| < $ 75 cm were accepted for further analysis. The event vertex was not included in primary-track $p_t$ fits. That procedure eliminated pileup-event tracks, selected those events well-positioned relative to the TPC and minimized correlations of individual track $p_t$ and $p_t$ spectrum shape with event multiplicity or event triggering not related to collision dynamics.

The resulting $p_t$ spectra were corrected for tracking efficiency, backgrounds and momentum resolution. Tracking acceptance and efficiency on $(p_t,\eta,z)$ and backgrounds were determined by embedding Hijing events in data events with at least one empty bunch (so-called abort-gap events). The same fractional correction was applied to all multiplicity classes. The correction factor  was 1.45 at 0.2 GeV/c, falling to 1.2 at 0.5 GeV/c and thereafter smoothly to 1 at 6 GeV/c. Efficiency- and acceptance-corrected (but not $p_t$-extrapolated) spectra integrate to multiplicity $n'_{ch} = (1.35\pm 0.015)\, \hat n_{ch}$, while the corrected {\em and} $p_t$-extrapolated per-event spectra integrate to `true' multiplicity $ n_{ch} = (2.0\pm0.02)\, \hat n_{ch}$. The errors reflect the spectrum-to-spectrum {\em relative} normalization uncertainties most relevant to this differential analysis. The normalization uncertainy common to all spectra is about 10\%.


 

\begin{figure}[h]
\includegraphics[keepaspectratio,width=3.3in]{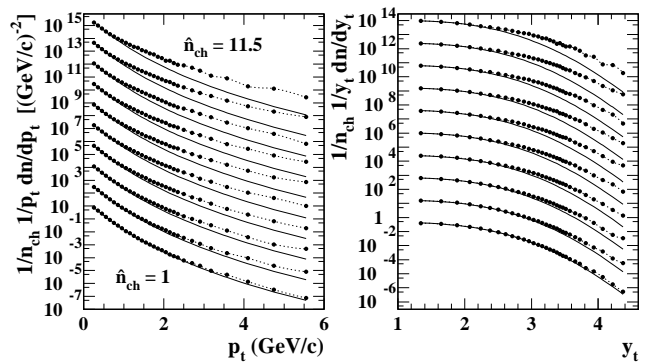}
\caption{\label{fig1}
Corrected and normalized charged-particle spectra on transverse momentum $p_t$ (left) and transverse rapidity $y_t$ (right) for 10 event multiplicity classes, displaced upward by successive factors 40 relative to $\hat n_{ch} = 1$ at bottom. Solid curves represent reference function $n_s/n_{ch}\cdot S_0(y_t)$ ({\em cf.} Sec.~\ref{s0def}). Dotted curves are spline fits to guide the eye.
}
\end{figure}

In Fig.~\ref{fig1} (left panel) corrected and normalized per-event $p_t$ spectra are plotted as points  in the form $1/n_{ch} \, 1/p_t\, dn/dp_t$ for ten multiplicity classes, offset by successive factors 40 (except for $\hat n_{ch} = 1$ at bottom). Parentheses for ratio prefactors of spectrum densities in the form $dn/dx$ are omitted to lighten notation. In other cases ratio prefactors are separated from densities by a dot. Corrected and extrapolated spectra normalized by $n_{ch}$ all integrate to unity in the sense of Eq.~(\ref{runav}) for $p_t$ or $y_t$, with the integration limit $\rightarrow \infty$. In Fig.~\ref{fig1} (right panel) equivalent spectra on {transverse rapidity} are plotted. Hard parton scattering leading to transverse fragmentation may be better described on transverse rapidity $y_t = \ln\left\{(m_t + p_t) / m_0  \right\}$, with transverse mass $m_t \equiv \sqrt{p_t^2 + m_0^2}$ and pion mass $m_\pi$ assumed for $m_0$: $y_t =2 \Rightarrow p_t \sim 0.5$ GeV/c and $y_t =4.5 \Rightarrow p_t \sim 6$ GeV/c.  The solid curves $n_s / n_{ch}\cdot S_0$ provide a visual reference for the data. $n_s(\hat n_{ch})$ and $S_0(p_t\text{ or } y_t)$ are defined below, and function $S_0$ is by definition independent of $\hat n_{ch}$.

\section{Power-law analysis}

The {\em power-law} function is the conventional model function applied to $p_t$ spectra from relativistic nuclear collisions~\cite{ua1}. Said to be `QCD-inspired,' the function $A/(1+p_t / p_0)^n$ goes asymptotically to $p_t^{-n}$  at large $p_t$ (hence `power-law') and approximates an exponential at small $p_t$. The argument supporting the power-law function assumes that $p_t$ spectra at larger collision energies can be modeled with a single functional form. In this part of the analysis we test that assumption.  The $p_t$ spectra for ten multiplicity classes in Fig.~\ref{fig1} were fitted with the three-parameter power-law model function defined above. Parameters $A$, $p_0$ and $n$ were independently varied to minimize $\chi^2$ for each multiplicity class (in all fitting $\chi^2$ was calculated using only statistical errors). The inclusive mean $p_t$ was extracted for each class as $\langle p_t \rangle \equiv 2 p_0 / (n-3)$ ({\em cf.} Sec.~\ref{meanpt} for those results). 

\begin{figure}[h]
\includegraphics[height=1.65in,width=3.3in]{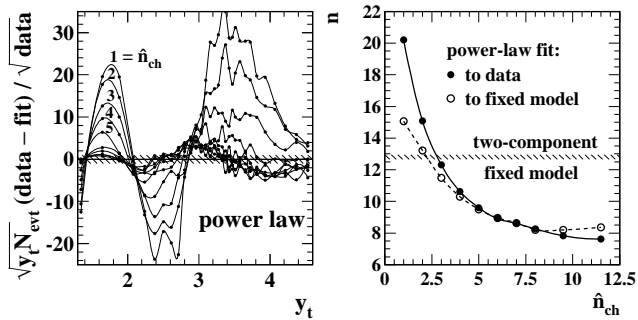}
\caption{\label{fig1a}
Left: Relative residuals from power-law fits to $p_t$ spectra in Fig.~\ref{fig1}.  The hatched band represents the expected statistical errors for STAR data. Right: Exponents $n$ from power-law fits to data (solid points) and to corresponding two-component fixed-model functions (open circles, see Sec.~\ref{twocomp1}) compared to the two-component fixed-model L\'evy exponent $12.8\pm0.15$ (hatched band). 
}
\end{figure}


In Fig.~\ref{fig1a} (left panel) we plot {\em relative} fit residuals $\sqrt{y_t\, N_{evt}}$ (data $-$ fit) / $\sqrt{\text {data}}$ distributed on $y_t$. The points indicate the actual data positions. The quantity plotted insures that the residuals are directly measured in units of the {\em r.m.s.} statistical error at each $y_t$. These relative residuals are then similar to Pearson's correlation coefficient or relative covariance~\cite{pearson}. Poisson errors apply to $dn/dy_t$, whereas the spectra plotted in Fig.~\ref{fig1} (`data') are of the form $1/y_t \,\, dn/dy_t$. Thus, a factor $\sqrt{y_t}$ is required to make the statistical reference uniform on $y_t$ in these residuals plots. The residuals structure on $p_t$ is equivalent to that on $y_t$ within a Jacobian factor (the fits were actually done on $p_t$ and the residuals transformed to $y_t$ for this plot). As noted in the discussion of Fig.~\ref{fig4} and elsewhere, much of the structure due to hard scattering and fragmentation is displaced to small $p_t$ in a nonlinear way when plotted on $p_t$.

The large-wavelength residuals in Fig.~\ref{fig1a} (left panel) exceed the expected statistical error (hatched band) by up to $30\times$ and are similar in form for various $\hat n_{ch}$ classes, revealing a large systematic disagreement between the power-law model and data. The small-wavelength structure, mainly attributable to true statistical fluctuations, is consistent with expectations (hatched band).  The argument supporting the power-law model of $p_t$ spectra is thus shown to fail when tested with high-statistics STAR p-p data. 

In Fig.~\ref{fig1a} (right panel) we plot best-fit values of power-law exponent $n$ {\em vs} $\hat n_{ch}$ resulting from fits to data (solid points) and to the two-component model functions described later in this paper (open circles). The latter points and hatched band are discussed in Sec.~\ref{discussion}. We observe a very strong variation of $n$ with multiplicity. Reduction of $n$ with increasing hard scattering is expected in the power-law context, but we find that the physical mechanism is different from the theoretical expectation ({\em cf.} Sec.~\ref{discussion}).

{
We observe very strong disagreement between the power-law model function and data, whereas a previous UA1 (SP\=PS) analysis reported power-law fits with reasonable $\chi^2$ at the same energy~\cite{ua1}. The UA1 results are nevertheless {consistent with} the present analysis because that analysis was inclusive on $n_{ch}$ and employed only 20k minimum-bias events ({\em vs} $3 \times10^6$ for the present analysis). That analysis was therefore statistically insensitive to the structures apparent in Fig.~\ref{fig1a}.
} 
Statistics for the UA1 minimum-bias $p_t$ spectrum are comparable to the $\hat n_{ch} = 11.5$ multiplicity class in this study, but the latter contains about 10$\times$ the hard component in the UA1 minbias spectrum. An E735 (FNAL) analysis of spectrometer data at 0.3, 0.55, 1.0 and 1.8 TeV~\cite{e735}, including multiplicity dependence of spectrum shapes, also obtained satisfactory power-law fits to $p_t$ spectra. However the effective event number was comparable to the UA1 study, in part because of the reduced angular acceptance of the spectrometer relative to the STAR CTB detector, and the $p_t$ acceptance [0.15,3] GeV/c was considerably less than STAR or UA1, further reducing sensitivity to spectrum shape. Given this exclusion of the power-law model we now seek an alternative  model which best describes $p_t$ spectra from relativistic nuclear collisions.

\section{Running Integration}



Running integration provides substantial noise reduction for spectrum analysis, thereby improving precision. In this section we examine the $n_{ch}$ dependence of differential and integrated spectra and define alternative normalization factor $n_s(\hat n_{ch})$ and reference function $S_0$. 

\subsection{Spectrum normalization} \label{specnorm}

\begin{figure}[h]
\includegraphics[width=1.65in,height=1.65in]{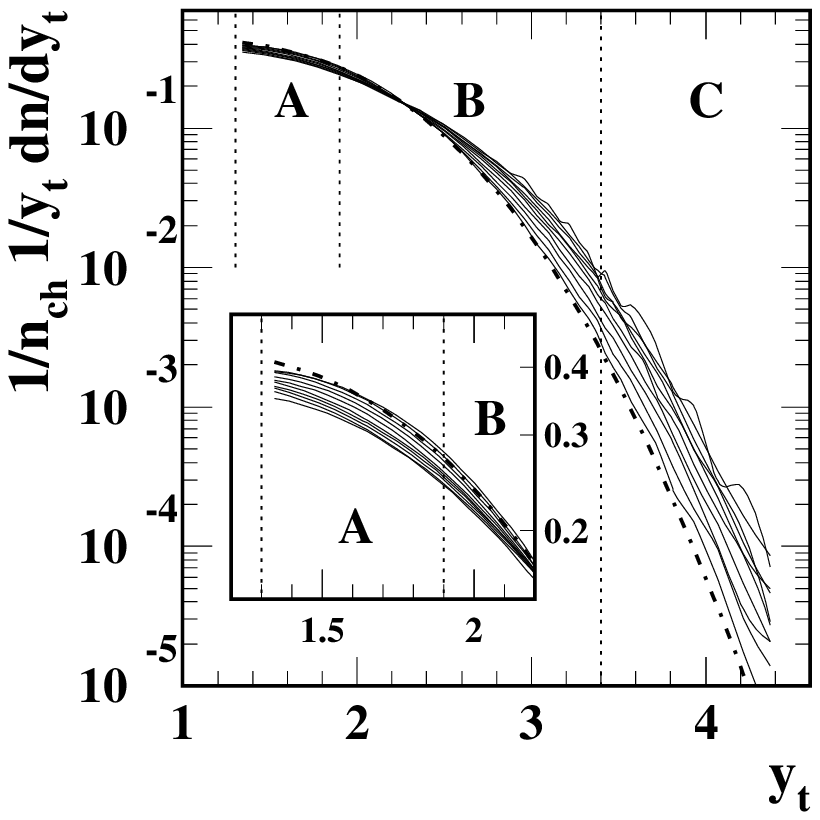}
\includegraphics[width=1.65in,height=1.65in]{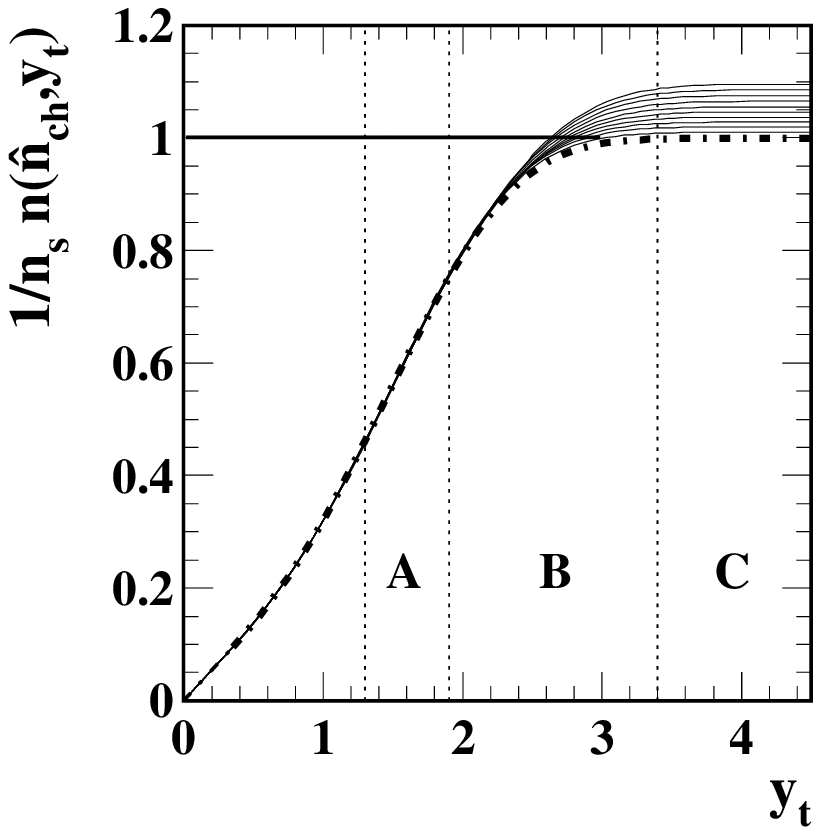}
\caption{\label{fig2a}
Left: Spectra from Fig.~\ref{fig1} (right panel) replotted as spline curves and without offsets (solid curves) compared to reference $S_0$ (dash-dot curve).
Right: Running integrals Eq.~(\ref{runav}) of extrapolated $y_t$ spectra in Fig.~\ref{fig1} divided by $n_s / n_{ch}$ (solid curves) compared to running integral $N_0(y_t)$ of reference $S_0(y_t)$ (dash-dot curve). The ten data curves from bottom to top correspond to increasing $\hat n_{ch} \in [1,11.5]$. 
}
\end{figure}

In Fig.~\ref{fig2a} (left panel)  the spectra from Fig.~\ref{fig1} (right panel) are replotted without vertical offsets as spline curves for detailed comparison. No assumptions have been made about the data, and all spectra integrate to unity when extrapolated. The dash-dot curve is reference  $ S_0$ defined in this section. To facilitate the discussion we identify three regions on $y_t$ separated by the vertical dotted lines:  A = [1.3,1.9], B = [1.9,3.4] and C = [3.4,4.5]. The region below $y_t = 1.3$ is outside the $p_t$ acceptance. Regions A and C are defined such that the curves within them are nearly constant relative to one another, whereas in region B the differences between curves vary rapidly. 

The trend of the spectra with increasing $\hat n_{ch}$ is counterbalancing changes  within A and C: linear decrease in A (see inset) and linear increase in C. The relative variation in the two regions over the observed $\hat n_{ch}$ range is quite different: 10\% reduction in A and $10\times$ increase in C. Such balancing variations are expected if the yield in C increases relative to A with $\hat n_{ch}$, due to the requirement that the normalized spectra must integrate to unity. We conclude that with increasing $\hat n_{ch}$ additional particle yield localized on $y_t$ and dominating region C is added to the spectrum. 

The apparent  reduction at smaller $y_t$ is then a trivial effect of the unit-integral condition which can be compensated by changing the normalization. We normalize the spectra not by true total multiplicity $n_{ch}$ but by multiplicity $n_s$ defined such that the normalized spectra approximately coincide within region A. The variation of lower end-point positions with $\hat n_{ch}$ is compensated within errors by normalizing with the linear function $\tilde n_s(\hat n_{ch}) =2 \hat  n_{ch} \,(1-0.013\, \hat n_{ch})$ (function $\tilde n_s$ estimates multiplicity $n_s$). The negative term compensates the relative yield increase at larger $y_t$. The revised normalization also facilitates the running integration study described below.

\subsection{Running integrals and reference $S_0$}

To calculate  running integrals the measured spectra are extrapolated in the $p_t$ interval [0,0.2] GeV/c ($y_t \in [0,1.15]$) with reference function $n_s / n_{ch}\cdot S_0$. The extrapolation is relatively  insensitive to the $S_0$ parameters, insuring quick convergence of the $S_0$ optimization procedure described below. The running integral of a $y_t$ spectrum is defined by
\bea \label{runav} 
n(\hat n_{ch},y_t) = \int_0^{y_t} \hspace{-.1in} dy'_t\, y'_t\, \left\{1/y'_t\, dn(\hat n_{ch},y'_t)/dy'_t\right\}.
\eea
In Fig.~\ref{fig2a} (right panel) the normalized running integrals $1/n_s(\hat n_{ch}) \cdot n(\hat n_{ch},y_t)$ reveal the detailed structure of the spectra with much-improved signal-to-noise ratio. We observe  that the integrals in the right panel  indeed nearly coincide up to $y_t \sim 2$. Above that point (region B) the integrals separate. In region C the integrals all saturate, with nearly equal spacings between curves. That result provides a first detailed look at the localized (on $y_t$) additional yield which produces the $n_{ch}$ dependence of the $y_t$ spectrum shape. 

Given the results in Fig.~\ref{fig2a} (right panel) the natural choice for a reference is one which coincides with all data curves for $y_t < 2$ and defines a limiting case for the sequence of separated data curves at larger $y_t$. We therefore {\em define} the reference as the asymptotic limit of the $y_t$ spectra (or their integrals) as $\hat n_{ch} \rightarrow 0$.  For reasons discussed below we chose as a trial reference the {\em L\'evy} distribution~\cite{wilke} 
\bea
S_0(m_t;\beta_0,n) = A_s/(1+\beta_0\, (m_t - m_0)/n)^n
\eea
defined on transverse mass $m_t$ and suitably transformed to $y_t$. $\beta_0 \equiv 1/T_0$ is an inverse-slope parameter. We find that the L\'evy distribution with optimized parameters (dash-dot reference curves in Fig.~\ref{fig2a}) coincides with the desired asymptotic form. Determination of parameters $n$ and $\beta_0$ from the data is discussed in the next subsection. Amplitude $A_s(\beta_0,n)$ is defined by the unit-integral normalization requirement on $S_0$.

The running integral of $S_0$, the dash-dot curve in Fig.~\ref{fig2a} (right panel) denoted by  $N_0$, is obtained by replacing the curly bracket in Eq.~(\ref{runav}) with $S_0(y_t)$, in which case $n(\hat n_{ch},y_t) \rightarrow N_0(y_t)$ (also, see the legend in Fig.~\ref{fig2} -- right panel). $N_0$ is thereby defined as the limit as $\hat n_{ch} \rightarrow 0$ of the running integrals for the ten multiplicity classes. We can obtain a more differential picture by optimizing reference curve $S_0$ and subtracting it and its running integral $N_0$ from the data. Fig.~\ref{fig2} (left panel) discussed in the next subsection reveals the $n_{ch}$-dependent yield increase as a localized structure on $y_t$ and is used to optimize $S_0$. This differential procedure represents a new level of precision in spectrum analysis facilitated by the high-statistics STAR p-p data and the running-integral technique.

\subsection{Optimizing reference $S_0$} \label{s0def}

\begin{figure}[h]
\includegraphics[width=1.65in,height=1.65in]{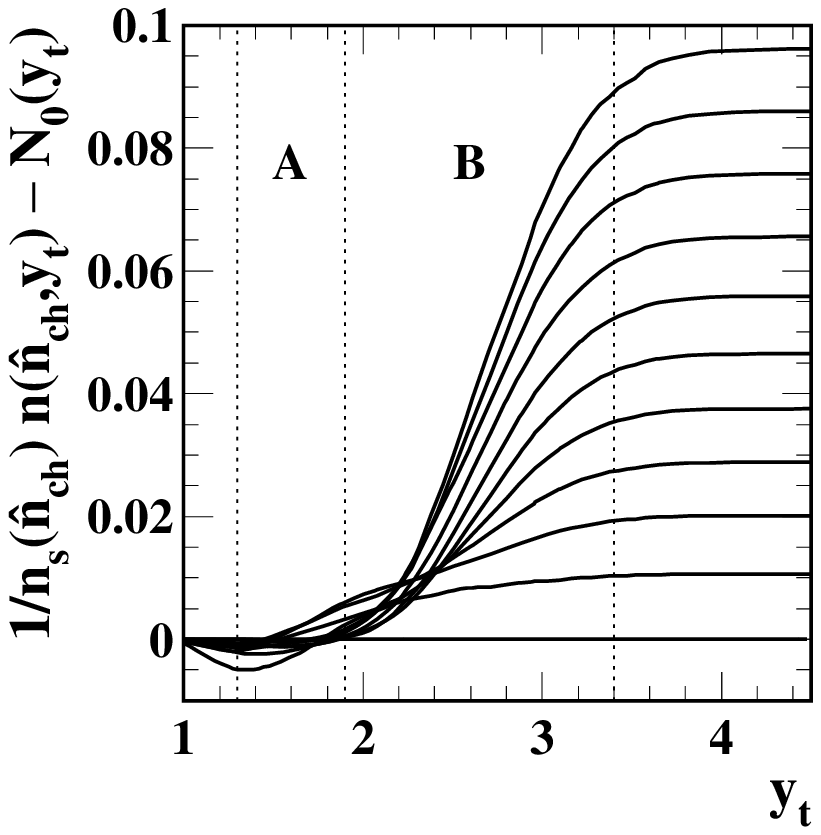}
\includegraphics[width=1.65in,height=1.65in]{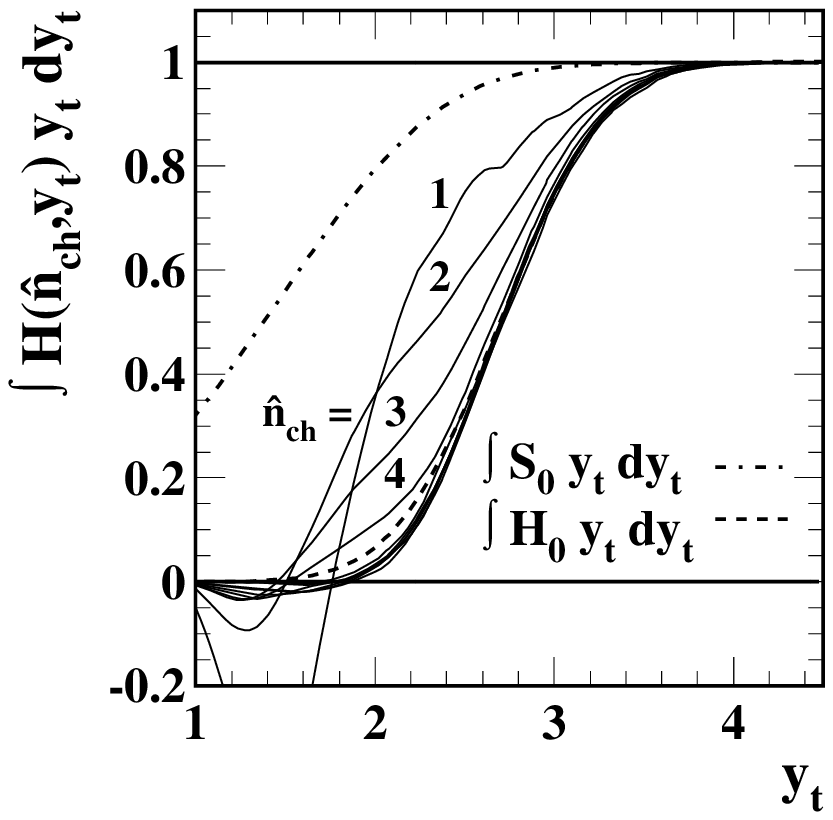}
\caption{\label{fig2}
Left: Differences between integrals of extrapolated $y_t$ distributions in Fig.~\ref{fig1} according to Eq.~(\ref{runav}), and integral $N_0(y_t)$ of soft reference $S_0(y_t)$. The ten curves correspond to $n_{ch} \in [1,11.5]$ from bottom to top. Right: Distributions in the left panel divided by their end-point values at $y_t = 4.5$. The dashed curve is the running integral of $H_0$ ({\em cf.} Sec.~\ref{diff}). The dash-dot curve is $N_0$, the running integral of $S_0$.
}
\end{figure}

In Fig.~\ref{fig2} (left panel) we plot the difference between running integrals $1/n_s(\hat n_{ch}) \cdot n(\hat n_{ch},y_t)$ of the corrected spectra in Fig.~\ref{fig2a} (right panel) and reference integral $N_0(y_t)$ (the dash-dot curve in that panel). In region B we observe a strong localized $\hat n_{ch}$ dependence in the $y_t$ spectra. The optimum parameters for $S_0$ are derived as follows. Inverse-slope parameter $\beta_0$ is adjusted to minimize residuals in region A of Fig.~\ref{fig2} (left panel). $\beta_0$ determines the {\em average slope} of the residuals in that region. Exponent $n$ then determines the size of the {\em first step} in region C. $n$ is adjusted so that the first step follows the nearly linear trend of $n_{ch}$ dependence in that $y_t$ interval. Amplitude $A_s(\beta,n)$ is determined by the unit-normalization requirement for $S_0$. 

Changing either $\beta_0$ or $n$ in $S_0$ {\em does not alter} the step-wise variation with $\hat n_{ch}$ of the data curves  in the left panel above the first step. That structure is inherent in the data and unaffected by the reference choice ({\em cf.} Fig.~\ref{fig2a} - right panel, before reference subtraction). The amplitude variation within region C is well represented by $n_h / n_s = \alpha\, \hat n_{ch}$ with $\alpha  \sim 0.01$, where $n_h$ is the coefficient of $H_0$ defined in the next subsection. That procedure determines reference $S_0$ parameters $A_s = 20.3\pm0.1,  n=12.8\pm0.15$ and $T_0 = 0.1445\pm0.001$ GeV~\cite{levy-n}. 


In Fig.~\ref{fig2} (right panel) the curves are obtained by dividing the curves in the left panel by their values at  upper endpoint $y_t = 4.5$ which approximate ratio $n_h / n_s$. Reference $N_0(y_t)$ is included in the right panel as the dash-dot curve. Comparing $N_0$ to the data integrals it is clear that the multiplicity dependence in Fig.~\ref{fig2} cannot be accommodated by adjusting $S_0$. With the exception of the first few $\hat n_{ch}$ values (labeled curves) the integrals closely follow a common trend: an error function or running integral of a gaussian which estimates {\em in a model-independent way} the running integral of the $n_{ch}$-independent model function $H_0(y_t)$ determined differentially in the next section.

\section{Differential Analysis} \label{diff}

Using running integrals we have defined a precision reference for the $y_t$ spectra and isolated the $n_{ch}$ dependence of those spectra relative to the reference.  We now return to the differential $y_t$ spectra and identify an additional spectrum component by subtracting the reference from the data. The dashed curve in Fig.~\ref{fig2} (right panel) (just visible near $y_t = 2$) represents the running integral of model function  $H_0$ determined in this section. $H_0(y_t)$ models the additional yield at larger $y_t$ as a differential $y_t$ spectrum component. It is already clear from Fig.~\ref{fig2} that the shape of that component is approximately gaussian and nearly independent of $n_{ch}$.

\begin{figure}[h]
\includegraphics[keepaspectratio,width=3.3in]{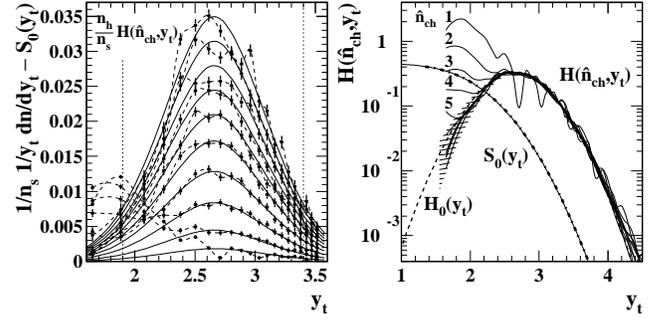}
\caption{\label{fig3}
Left panel: Distributions on $y_t$ in Fig.~\ref{fig1} (right panel) divided by $n_s/n_{ch}$ minus reference $S_0(y_t)$. Dashed curves in the left panel and solid curves in the right panel are spline fits to guide the eye. The vertical dotted lines enclose region B previously defined. Right panel: Distributions $H(n_{ch},y_t)$ obtained by dividing the curves in the left panel by $n_h/n_s$. The dashed curve represents hard reference $H_0(y_t)$. The dash-dot curve represents soft reference $S_0(y_t)$. The solid curve underlying the dash-dot curve is an error function~\cite{levy-err}. The hatched region estimates the systematic error from subtraction of $S_0$.  
} 
\end{figure}

In Fig.~\ref{fig3} (left panel) we show the result of subtracting reference $S_0(y_t)$ from the $y_t$ spectra in Fig.~\ref{fig1} (right panel) divided by $n_s / n_{ch}$. We obtain the difference distributions denoted by $n_h / n_{s} \cdot H(\hat n_{ch},y_t)$ (the data points connected with dashed curves). Those data represent {\em all} $n_{ch}$ dependence of the $y_t$ spectra relative to fixed reference $S_0$. The error bars denote statistical errors, applicable also to the data in Fig.~\ref{fig1}. The two vertical dotted lines enclose region B on $y_t$ previously defined. $H(\hat n_{ch},y_t)$ has unit integral by definition, consistent with $n_s + n_h = n_{ch}$. The shapes of the data curves are well-approximated by the unit-integral gaussian reference 
\bea
H_0(y_t;\bar y_t,\sigma_{y_t}) = A_h(\bar y_t,\sigma_{y_t}) \cdot \exp\left\{- \frac{1}{2} \left[\frac{y_t - \bar y_t}{\sigma_{y_t}}\right]^2 \right\},
\eea 
with $A_h = 0.335\pm0.005,\, \bar y_t = 2.66 \pm 0.02$ and $\sigma_{y_t} = 0.445 \pm 0.005$. The solid curves represent  $n_h / n_s\cdot H_0$, with best-fit amplitudes $n_h(\hat n_{ch}) / n_s(\hat n_{ch})$ plotted in Fig.~\ref{fig4a} (right panel, solid dots). $n_h$ is the multiplicity of the new spectrum component. The data are generally well described by the model, except for the excursions at smaller $y_t$ for the smaller $\hat n_{ch}$ values.



Dividing the data in Fig.~\ref{fig3} (left panel) by the corresponding best-fit gaussian amplitudes $n_h / n_s$ reveals the normalized data distributions $H(\hat n_{ch},y_t)$ in the right panel. Reference $S_0(y_t)$, shown as the dash-dot curve in the right panel, is approximately an error function~\cite{levy-err}.  The hatched region estimates the systematic error from the $S_0$ subtraction.  Deviations from the $H_0$ model function (dashed curve) in that panel represent {\em all} the residual $n_{ch}$ dependence of the $y_t$ spectra, {\em i.e.,} all deviations from the two-component model in Eq.~(\ref{2compalg}) below. Those deviations  are plotted in Fig.~\ref{fig3a} (left panel) and discussed further in the next section.

The QCD-based power-law trend $p_t^{-n}$ expected for hard parton scattering would appear in this plotting format as a straight line with negative slope equal to the exponent or `power' $-n$~\cite{mjtann}, since $y_t \sim \ln(2p_t/m_0)$ at large $p_t$ makes the plot effectively a log-log plot. Out to $y_t = 4.5$ or $p_t = 6$ GeV/c we observe no linear tangential departure from gaussian model $H_0$ (dashed parabola) in data $H(\hat n_{ch},y_t)$.

\section{Two-component model} \label{twocomp1}


The two-component model~\cite{tcomp1,duke} states that the minimum-bias frequency distribution on event multiplicity from relativistic p-p collisions can be resolved into two components, each approximated by a negative binomial distribution (NBD) with its own mean and $k$ parameter. The two components correspond to events with (hard) and without (soft) significant hard parton scatters. That concept can be extended to the possibility that the inclusive $p_t$ spectrum shape for hard events is different from that for soft events~\cite{cdf}---that the former contains an additional {\em spectrum} component which we designate the hard component, the complement being then the soft spectrum component. In that interpretation spectra from different multiplicity classes should contain different admixtures of the two spectrum components, and the multiplicity dependence of the spectrum shape may therefore provide a means to isolate those components. 

{
In  this section we examine the two-component model in detail. We consider the factorization structure of the model function that has emerged from data analysis, we examine the residuals structure compared to statistical errors and then test the {\em necessity} of the fixed-parameter model function by fitting the data with all model parameters freely varying. We finally relate all multiplicities in the model and show that they form a consistent system.
}


\subsection{Two-component model function}

We have analyzed the multiplicity dependence of $y_t$ spectra from p-p collisions without an {\em a priori} model and have observed a strong $n_{ch}$ dependence whose functional forms we now summarize. The two-component model of $y_t$ spectrum structure can be generally represented by the first line of
\bea \label{2compalg}
1/y_t\, dn / dy_t &=& s(\hat n_{ch},y_t) + h(\hat n_{ch},y_t) \\ \nonumber
& = & n_s(\hat n_{ch})\, S_0(y_t) + n_h(\hat n_{ch})\, H_0(y_t) + \dots,
\eea
with unspecified soft and hard spectrum components $s(\hat n_{ch},y_t)$ and $h(\hat n_{ch},y_t)$. What we have inferred from the $n_{ch}$ dependence of the measured $y_t$ spectra is the second line, which represents a factorization hypothesis with spectrum components modeled by unit-normal functions $S_0(y_t)$ and $H_0(y_t)$ independent of $n_{ch}$, ratio $n_h(\hat n_{ch}) / n_{s}(\hat n_{ch}) = \alpha\, \hat n_{ch}$, and constraint $n_s + n_h= n_{ch}$. We suggest that the algebraic model in the second line corresponds to the two-component physical model described above and represented by the first line.
In the rest of this section we consider the quality and details of the parameterized model in Eq.~(\ref{2compalg}) and test its uniqueness by performing a free $\chi^2$ fit of the unconstrained model functions to the data.

In the power-law context there is no {\em a priori} hypothesis for $n_{ch}$ dependence: each of the ten multiplicity classes in this analysis can be fitted independently with the three-parameter model to produce 30 fit parameters. The corresponding residuals are shown in Fig.~\ref{fig1a} (left panel). For the two-component model we could in principle have six free parameters for each $\hat n_{ch}$, producing 60 fit parameters. However, the algebraic model of Eq.~(\ref{2compalg}) (second line) contains constraints motivated by the requirement of model simplicity which greatly reduce the number of independent parameters. 1) The shapes of unit-integral functions $S_0(y_t)$ and $H_0(y_t)$ are independent of multiplicity: each function is determined by only two parameters fixed for all $n_{ch}$. 2) The relative normalization of the two components is nearly linearly proportional to the observed multiplicity, as defined by fifth parameter $\alpha$. Thus, only five parameters represent all the data in that model. As with the power-law model we compare data to model on the basis of {\em relative} fit residuals on $y_t$, which provide a more differential and direct assessment of fit quality than the $\chi^2$ statistic.

\subsection{Five-parameter fixed model}

\begin{figure}[h]
\includegraphics[width=1.65in,height=1.65in]{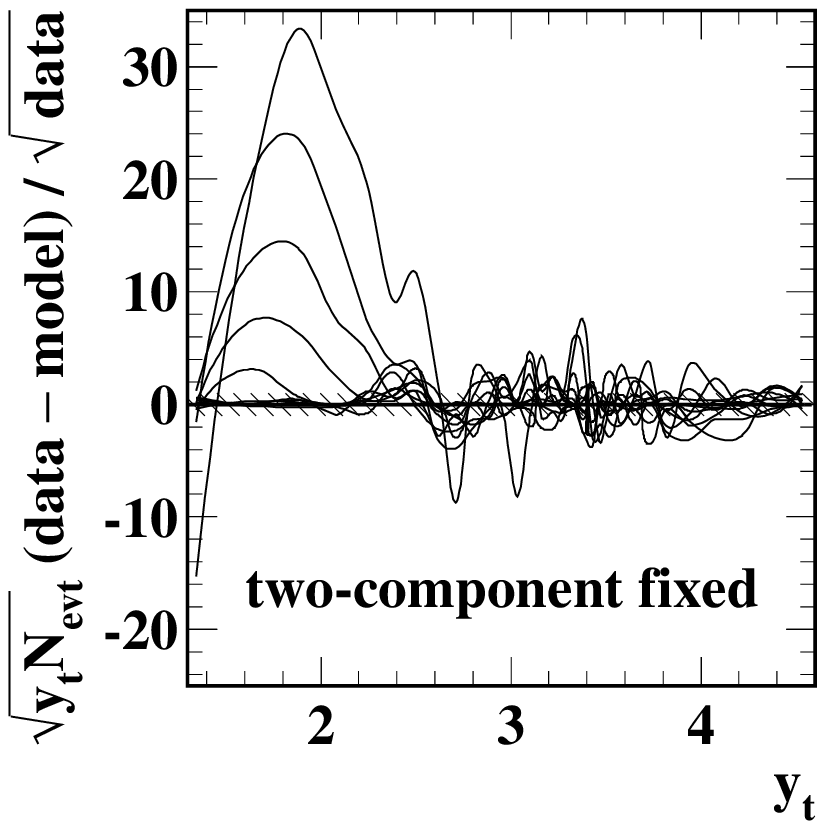}
\includegraphics[width=1.65in,height=1.65in]{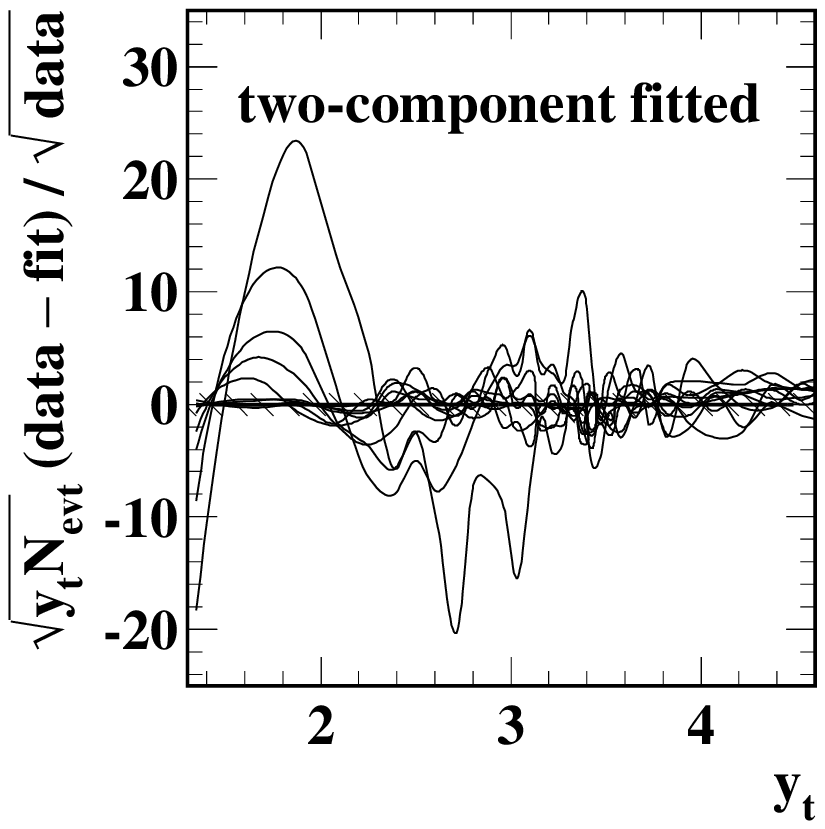}
\caption{\label{fig3a}
Left: Relative residuals between data $y_t$ spectra in Fig.~\ref{fig1} (right panel)  and the two-component fixed parameterization, for all multiplicity classes.
Right: Relative residuals between data $y_t$ spectra and two-component free $\chi^2$ fits, also for all multiplicity classes.  The hatched bands represent the expected statistical errors for STAR data.
} 
\end{figure}

The residuals in Fig.~\ref{fig3a} (left panel) correspond to the function in the second line of Eq.~(\ref{2compalg}) with five optimized parameters held fixed for all $\hat n_{ch}$. Above $y_t = 2.7$ the residuals are consistent with statistical fluctuations except for a few sharp structures with amplitude several times the statistical error. Those structures arise from the comparatively low statistics of the Monte Carlo simulations used for background corrections. The Monte Carlo statistical fluctuations appear in these residuals as small-wavelength systematic deviations. 

The prominent residuals in $y_t < 2.7$ for $\hat n_{ch} =$ 1-4 (a `third component') could represent nontrivial $n_{ch}$ dependence of the soft or hard component or some additional physical mechanism. The endpoint values at $y_t = 4.5$ in Fig.~\ref{fig2} (left panel) vary linearly with $n_{ch}$ to a few percent (open symbols in Fig.~\ref{fig4a} -- right panel), despite the substantial nonlinear excursions at small $y_t$ of the distributions in Figs.~\ref{fig3} and~\ref{fig3a}. That {\em apparent} contradiction suggests that the prominent residuals may represent a change of the hard component at small $n_{ch}$ which preserves the linear trend of the integrals. These two-component residuals from the five-parameter fixed model are otherwise {\em much smaller} than the systematic deviations of the power-law model in Fig.~\ref{fig1a} (left panel) with its 30-parameter $\chi^2$ fit, especially in the large-$y_t$ region where the power-law model should be most applicable. 



\subsection{Two-component free $\chi^2$ fits} \label{chi2}

To determine whether the algebraic model of Eq.~(\ref{2compalg}) is {\em necessary} (required by the data), not simply an accident of data manipulation, spectra for $\hat n_{ch} \in [1,11.5]$ were fitted with the six-parameter function in Eq.~(\ref{2compalg}) using $\chi^2$ minimization. Spectra $ 1/y_t\, dn/dy_t$ were first normalized by multiplicity estimator $\tilde n_s(\hat n_{ch})$ from the fixed parameterization. The coefficients of $S_0$ and $H_0$ in the fitting function are then $n_s / \tilde n_s$ and $n_h / \tilde n_s$. The six parameters $(n_s,\beta_0,n,n_h,\bar y_t,\sigma_{y_t})$ were freely varied for each $\hat n_{ch}$.

The residuals from the free fits are shown in Fig.~\ref{fig3a} (right panel). 
The fit residuals are comparable to the corresponding fixed-model residuals in Fig.~\ref{fig3a} (left panel), even though the free fits include six independent parameters for each of ten $n_{ch}$ classes for a total of 60 parameters, compared to the fixed model with only five parameters to describe all ten $n_{ch}$ classes. The residuals for the smaller $\hat n_{ch}$ values show that the free fit attempts to minimize the small-$y_t$ structure (`third component') in the left panel at the expense of increased intermediate-$y_t$ residuals. The effect on the fit parameters is however modest, as illustrated in Table~\ref{tab1}.

\begin{table}[h]
\begin{center}
\begin{tabular}{|c||c|c|c||c|c|c||c|} \hline
\multicolumn{1}{|c||}{ fitted} &
\multicolumn{3}{|c||}{soft component} &
\multicolumn{3}{|c||}{hard component} &
\multicolumn{1}{|c|}{ }\\ \hline
 & & & & & &   \vspace{-.13in} \\
 $\hat n_{ch}$ & $n_s/ \tilde n_{s}$ & $T_0$(GeV) & $n$ & $n_h/ \tilde n_{s}$ & $\bar y_t$ & $\sigma_{y_t}$ & $\chi^2 / \nu$  \\ \hline\hline
 1 & 0.995 & 0.145 & 11.97 & 0.000 & -- & -- & 73.3  \\ \hline
 2 & 1.001 & 0.145 & 11.78  & 0.002 & 2.75 & 0.500 & 40.4 \\ \hline
 3 & 1.001 & 0.145 & 11.83 & 0.013 & 2.75  & 0.421 & 15.0 \\ \hline
 4 & 0.996 & 0.145 & 11.74 & 0.025 & 2.75  & 0.400 & 7.36 \\ \hline \hline
 5 & 0.994 & 0.145 & 12.60 & 0.049 & 2.65 & 0.427 &  3.14 \\ \hline \hline
 6 & 1.001 & 0.144 & 15.63  & 0.089 & 2.57 & 0.450 & 1.09 \\ \hline
 7 & 0.999 & 0.144 & 15.42 & 0.097 & 2.57  & 0.451 & 0.62 \\ \hline
 8 & 1.005 & 0.144 & 16.73 & 0.115 & 2.56  & 0.454 & 1.18 \\ \hline
 9.5 & 1.011 & 0.143 & 16.66 & 0.130 & 2.56  & 0.456 & 0.52 \\ \hline
 11.5 & 0.995 & 0.145 & 15.69  & 0.128 & 2.58 & 0.460 & 1.20  \\ \hline \hline
& & & & & &   \vspace{-.13in} \\ 
 fixed & 1.000 & 0.1445 & 12.8  & $0.0105 \hat n_{ch}$ & 2.66 & 0.445  &  \\ \hline
 & & & & & &   \vspace{-.13in} \\
error & 0.005 & 0.001 & 0.15  & $0.0005 \hat n_{ch}$ & 0.02 & 0.005  &  \\ \hline
\end{tabular}
\end{center}
\caption{Two-component $\chi^2$ fit parameters. The line labeled `fixed' contains the two-component fixed-model parameters. Each fit has $\nu = 28$ degrees of freedom. The error row applies only to the fixed parameterization. The fit errors are generally smaller than those errors for the last five free-fit rows. Significant systematic effects are discussed in the text.
\label{tab1}}
\end{table}


Table~\ref{tab1} compares the fixed-model parameter values (fixed) to the results of the six-parameter free fits (fitted) for ten $\hat n_{ch}$ classes. If the hard-component gaussian on $y_t$ were not necessary we would expect the $\chi^2$ fit to converge to the soft-component L\'evy distribution as a proxy for the power-law function. The results in Table~\ref{tab1} indicate that most of the free-fit $S_0$ and $H_0$ shape parameters remain nearly constant within errors across the full $\hat n_{ch}$ interval. The hard-component gaussian amplitudes are definitely nonzero and monotonically increasing, consistent with the trends in Fig.~\ref{fig3} (left panel) obtained by subtracting $S_0(y_t)$ from the normalized spectra in Fig.~\ref{fig1} (right panel). 


\begin{figure}[t]
\includegraphics[width=1.65in,height=1.65in]{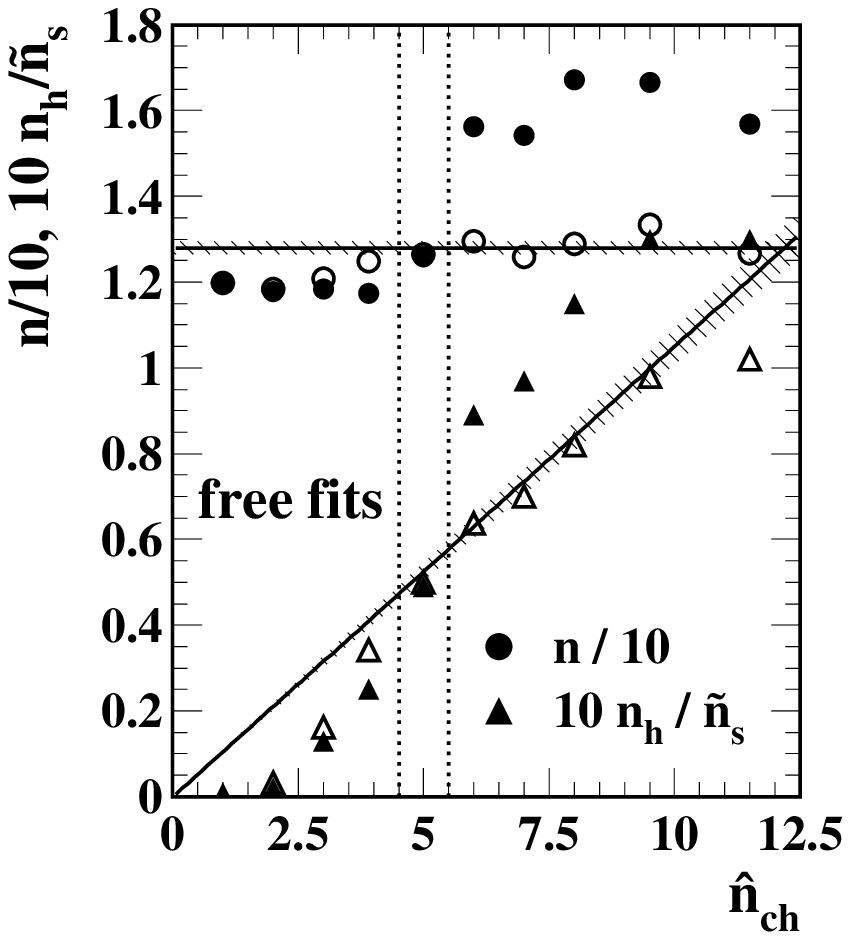}
\includegraphics[width=1.65in,height=1.65in]{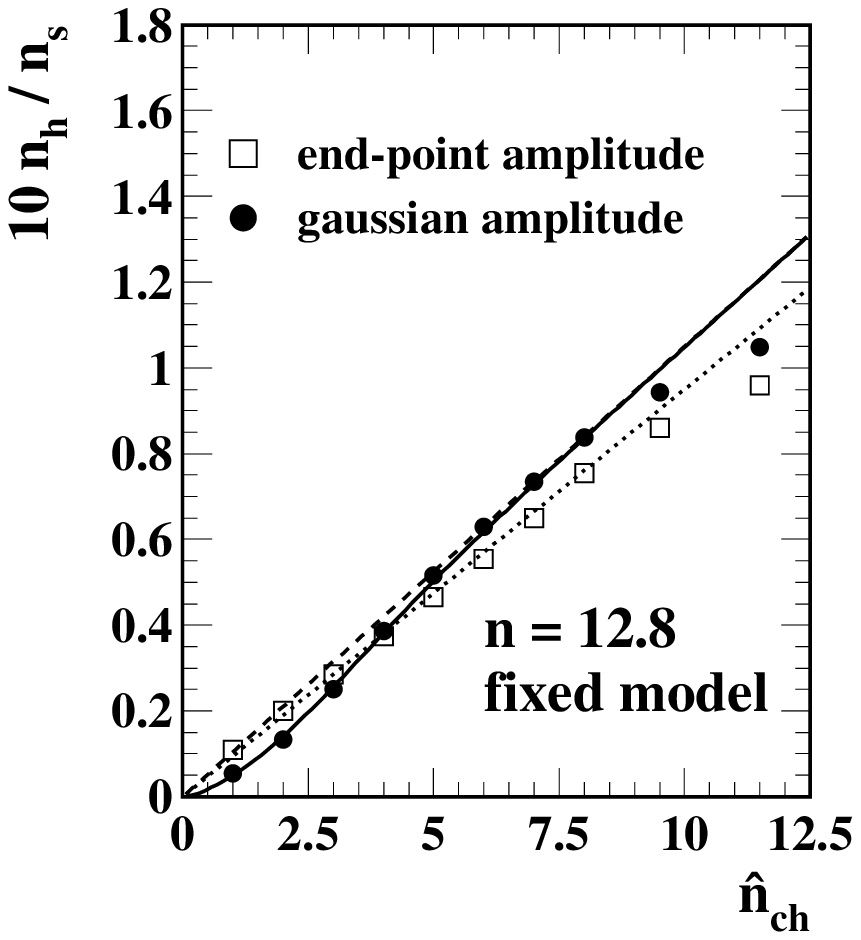}
\caption{\label{fig4a}
Left:   Parameters $n$ and $n_h / \tilde n_s$ {\em vs} $\hat n_{ch}$ from free $\chi^2$ fits (solid symbols) in Table I. The open symbols represent similar free fits but with $\bar y_{t}$ held fixed at 2.65 (see text for discussion). The bands represent the corresponding two-component fixed parameterizations with their stated errors also given in Table I.
Right: Ratio $n_h(\hat n_{ch}) / n_s(\hat n_{ch})$ derived from integrals in Fig.\ref{fig2} (left panel) (open squares) and from gaussian amplitudes in Fig.~\ref{fig3} (left panel) (solid dots). The dashed and dotted lines have slopes 0.105 and 0.095 respectively. The solid curve is described in the text.
} 
\end{figure}




Fig.~\ref{fig4a} (left panel) shows trends for the two fit parameters $n$ and $n_h / \tilde n_s$ which best illustrate the trade-off between soft/power-law and hard components of the model and the necessity of the two-component model. Best-fit values are presented for {all} $\hat n_{ch}$ classes as the solid symbols (open symbols are discussed below).  There are significant systematic deviations of exponent $n$ from the fixed-model value (hatched band) which are however qualitatively different from the trends in Fig.~\ref{fig1a} (right panel). The hard-component amplitude $n_h / \tilde n_s$ also deviates from the linear fixed-model trend, but the trend of monotonic increase is even stronger. The hard component appears to be {\em more} favored by the free fit than by the fixed parameterization. We discuss the systematic differences between fixed model and free fit in the following paragraphs. However, this fitting exercise does demonstrate that for almost all $\hat n_{ch}$ a two-component model is indeed {necessary} to describe RHIC p-p data.



The systematic deviations between free fits and fixed model in Fig.~\ref{fig4a} (left panel) are easily understood. We separately consider $\hat n_{ch} < 5$ and $\hat n_{ch} > 5$ (separated by the dotted lines). Generally, there is a strong positive correlation between soft-component exponent $n$ and hard-component relative amplitude $n_h/n_s$ originating from the requirement to describe the large-$y_t$ yield. If $n$ decreases the L\'evy distribution tail rises and the amplitude of the hard-component gaussian amplitude must decrease as well to compensate at large $y_t$, and conversely. The systematic deviations relative to the fixed model for $\hat n_{ch} < 5$ respond to the presence of the `third component,' which is not a part of the two-component model. To compensate for the additional component in the data the hard-component amplitude is suppressed and $n$ is reduced by about 10\% to provide additional yield from $S_0$ at small $y_t$. The consequence is negative residuals near $y_t = 2.6$ in Fig.~\ref{fig3a} (right panel).

For $\hat n_{ch} > 5$ a different issue arises. In Fig.~\ref{fig3} (left panel) we have noted previously that the hard-component data peaks are skewed (fall off more rapidly on the low-$y_t$ side) whereas the hard-component model function is a symmetric gaussian. The difference is most apparent in the running integrals of Fig.~\ref{fig2} (right panel): the dashed model curve lies above the data near $y_t \sim 2$. In Fig.~\ref{fig3} (right panel) the hatched region illustrates the region of maximum influence of the $S_0$ subtraction on the hard component. Because the hard-component data peaks are asymmetric the $S_0$ subtraction at larger $y_t$ must be reduced by increasing exponent $n$ (the small-$y_t$ $S_0$ contribution must remain constant to describe the spectra there). This requires a compensating increase in the hard-component amplitude to fit the larger-$y_t$ part of the spectra, and the gaussian model function must {\em shift down} on $y_t$ (by $\sim 0.1$ or 5 sigma) and the width increase slightly (0.01 or 2 sigma) to accommodate the {\em apparent} increased symmetry of the data hard component. 

To test that description the free fits were redone with the gaussian centroid fixed at $\bar y_t = 2.65$. The open symbols in Fig.~\ref{fig4a} (left panel) show the result. The best-fit parameters are now within the error bands of the fixed model, with only modest increase in $\chi^2/\nu$ (1.69, 1.07, 1.43, 0.95, 1.18 respectively for $\hat n_{ch} = 6,\cdots,11.5$ compared to the corresponding values in Table~\ref{tab1}). The fit residuals in Fig.~\ref{fig3a} (right panel) appear identical for the two cases. We emphasize that the mode (most probable point) of the data hard-component peak is near $y_t = 2.65$. The downward shift of the {\em model} peak in the free fit is a consequence of the skewness in the data hard component not described by the fixed model but consistent with measured fragmentation functions from reconstructed jets.

\subsection{Two-component multiplicities}    \label{mults}


In Sec.~\ref{specnorm} we adopted a normalization strategy which brought all spectra into coincidence in region A of Fig.~\ref{fig2a} (left panel) by defining multiplicity $n_s \propto \hat n_{ch}$ except for a small deviation linear in $\hat n_{ch}$. We then defined reference function $S_0$ as a limiting case of the spectrum $n_{ch}$ dependence and isolated a second component $H_0$ by subtracting the fixed reference from all spectra. The amplitude of $H_0$ relative to the reference is defined by ratio $n_h / n_s \propto \hat n_{ch}$. The representation to that point is (physics) model independent, derived only from the observed spectrum $\hat n_{ch}$ dependence: the reference is $\propto \hat n_{ch}$ and the second component is  $\propto \hat n^2_{ch}$. That difference is the {\em underlying basis} for distinguishing the two components.

In this section we have identified the two algebraic spectrum components with the components of a physical model of soft and hard parton scattering and subsequent fragmentation to detected particles. We distinguish four event multiplicities: 1) the observed multiplicity $\hat n_{ch}$ or uncorrected number of particles with $p_t > 0.2$ GeV/c in the STAR angular acceptance which serves as an event-class index, 2) the corrected and $p_t$-extrapolated multiplicity $n_{ch}$, 3) the `soft-component' multiplicity $n_s$ and 4) the `hard component' multiplicity $n_h$, with $n_s + n_h = n_{ch}$. We now examine the self-consistency of the multiplicities in our two-component model in the context of real spectrum properties, including efficiencies and acceptances.



Soft multiplicity $n_s$ is estimated by function $\tilde n_s(\hat n_{ch}) = [2.0\pm0.02\text{(rel)}\pm0.2\text{(abs)}]\, \hat n_{ch}\, [1 - (0.013\pm0.0005)\, \hat n_{ch}]$. The 1\% error applies to the relative or spectrum-to-spectrum normalization relevant to this differential analysis, whereas the 10\% error applies to the common normalization of all spectra. As noted, coefficient 0.013 is determined by requiring that corrected spectra normalized by $\tilde n_s$ approximately coincide within region A of Fig.~\ref{fig2a} (left panel) for all $\hat n_{ch}$. The factor 2 is determined by requiring that after correction, extrapolation with $S_0$ to $p_t = 0$ and normalization with $n_{ch}$ all spectra in Fig.~\ref{fig1} integrate to unity. In the first column of Table~\ref{tab1} deviations of $n_s / \tilde n_s$ from unity are consistent with the 1\% error estimate.


The hard fraction $n_h / n_s = \alpha\, \hat n_{ch}$ is estimated by two methods. In the first method we determine the gaussian amplitudes required to fit the data distributions in Fig.~\ref{fig3} (left panel). Those amplitudes give the solid gaussian curves compared to data in that plot and are plotted as the solid points in  Fig.~\ref{fig4a} (right panel). The linear trend (dashed line) corresponds to slope  $\alpha = 0.0105\pm0.0005$. The solid curve passing precisely through the points is $n_h(\hat n_{ch}) / n_s(\hat n_{ch}) = \{(0.0105\, \hat n_{ch})^{-10} + (0.005\, \hat n_{ch}^{1.5})^{-10} \}^{-1/10}$, the errors on the coefficients being $\pm$0.0005. The nonlinearity of that curve is related to the non-gaussian small-$y_t$ structure for small values of $\hat n_{ch}$ (third component).


In the second method we note that the distributions in the left panel of Fig.~\ref{fig2} are running integrals of data distributions in Fig.~\ref{fig3} (left panel). The amplitudes of those integrals at end-point $y_t = 4.5$, plotted as open squares in Fig.~\ref{fig4a} (right panel), also estimate ratio $n_h / n_s$. They vary nearly linearly (dotted line) with slope $\alpha = 0.0095\pm0.0005$. Reduction of $\alpha$ from 0.0105 for the gaussian amplitudes to 0.0095 for the integral endpoints results from small deviations of the data peaks from the $H_0$ gaussian model at small $y_t$ evident in Fig.~\ref{fig3}. The data are slightly skewed in a manner consistent with measured fragmentation functions. The model gaussians are matched to the data at and above the data peak mode or most probable point. The integral of any data peak is therefore expected to be slightly less than that of the corresponding model function. Both methods suggest saturation of the hard-component amplitude at larger $\hat n_{ch}$.



Consistency of the soft and hard multiplicity estimators within the two-component model can be established by the following argument: Tracking inefficiencies produce the same fractional changes for all $\hat n_{ch}$ and are represented by factors $\epsilon_s$ and $\epsilon_h$ for soft- and hard-component yields. The corrected spectra are extrapolated to $p_t = 0$ with soft model $S_0$. The fraction of $S_0$ falling above $p_t = 0.2$ GeV/c (within the $p_t$ acceptance) is represented by $\gamma$. The hard component identified in this analysis falls entirely within the $p_t$ acceptance. The {\em observed} multiplicity is then given by $\hat n_{ch} \equiv \gamma \epsilon_s\, n_s + \epsilon_h\, n_h$, whereas the corrected and extrapolated spectra integrate to true multiplicity $ n_{ch} = n_s + n_h$. 
The expression for $\hat  n_{ch}$ above can be rearranged to solve for $n_s$ in the first line below,
\bea
n_s &\simeq& \frac{\hat n_{ch}}{\gamma \epsilon_s} \, \left\{ 1 - \frac{\epsilon_h}{\gamma \epsilon_s} \cdot  \alpha\, \hat n_{ch} \right \} \text{~~~predicted} \\ \nonumber
\tilde n_s &=& 2 \hat n_{ch} \left\{ 1 -  0.013\, \hat n_{ch}   \right \} \text{~~~~observed}  
\eea
whereas the second line is the estimator inferred from the data. By integrating reference $S_0$ we determine that $\gamma = 0.7$: 70\% of the reference spectrum is within the acceptance $p_t > 0.2$ GeV/c. Tracking efficiencies $\epsilon_s$ and $\epsilon_h$ are both approximately 70\%, and we have determined from the data (running integrals) that $\alpha \sim 0.0095$. We therefore have $1/\gamma \epsilon_s \sim 2$ and  $\epsilon_h /\gamma \epsilon_s \cdot \alpha \sim 0.0135$, establishing the consistency (predicted $\leftrightarrow$ observed) of the two-component multiplicities. Coefficient 0.013 is identified as $\alpha / \gamma$, and the trend of $n_s$ is defined by ratio $n_h/n_s = \alpha\, \hat n_{ch}$. We thus close the circle, demonstrating quantitatively how increase with $\hat n_{ch}$ of the hard-component contribution to the spectrum forces $n_s$ to decrease relative to $\hat n_{ch}$ in compensation, why $\tilde n_s$ contains the negative term and what its magnitude must be.

\section{$\langle p_t \rangle$ Systematics} \label{meanpt}

Another aspect of the two-component model is the variation of $\langle p_t \rangle$ (inclusive mean $p_t$) with $\hat n_{ch}$. Estimation of $\langle p_t \rangle$ for spectra with incomplete $p_t$ acceptance requires either a model fit or direct integration of data with extrapolation. The power-law function for $p_t$ distributions ${1}/{p_t}\, dn / dp_t = {A}/{(1+p_t / p_0)^n}$, with $ \langle p_t \rangle = 2 p_0/ (n-3)$, has been used previously to extract $\langle p_t \rangle$ values from corrected $p_t$ spectra~\cite{ua1}. $\langle p_t \rangle$ can also be determined by direct integration of the experimental $p_t$ spectra, with extrapolation to $p_t = 0$ by a suitable model function. Finally, the two-component fixed-model function obtained in this analysis can provide a parameterization of $\langle p_t \rangle (n_{ch})$. 



The running multiplicity integral $n(\hat n_{ch},y_t) $ is defined by Eq.~(\ref{runav}), with the data extrapolated over $p_t \in [0,0.2]$ GeV/c by reference $n_s(n_{ch})\, S_0(p_t)$. Running integral $p_t(\hat n_{ch},y_t) $ can also be defined for transverse momentum $p_t$ by including an extra factor $p_t(y_t)$ in the integrand of Eq.~(\ref{runav}). The ratio $\langle p_t \rangle(\hat n_{ch},y_t) = p_t(\hat n_{ch},y_t)  / n(n_{ch},y_t)$ is then a function of $y_t$ for each value of $\hat n_{ch}$, and $\langle p_t \rangle(\hat  n_{ch})$ is the limit of that function as $y_t \rightarrow \infty$. $\langle p_t \rangle(\hat  n_{ch})$ is thus determined by direct integration of $p_t$ or $y_t$ spectra. 


A changing mixture of soft and hard components may cause $\langle p_t \rangle$ to vary with $n_{ch}$. The $\langle p_t \rangle$ values for individual components are obtained by direct integration of model functions $S_0$ and $H_0$: $\langle p_t \rangle_{soft} = 0.385\pm0.02$ GeV/c and $\langle p_t \rangle_{hard} = 1.18\pm0.01$ GeV/c.  A two-component {\em analytic expression} for $\langle p_t \rangle$ is then given by 
\bea \label{mnpt}
\langle p_t \rangle(\hat n_{ch}) = \left\{0.385 \frac{n_s(\hat n_{ch})}{n_{ch}} +  1.18 \frac{n_h(\hat n_{ch})}{n_{ch}}\right\} \text{GeV/c}, 
\eea
with $n_h /  n_{s} = \alpha \, \hat n_{ch}$ and $n_s + n_h = n_{ch}$. 



\begin{figure}[h]
\includegraphics[keepaspectratio,width=2.8in]{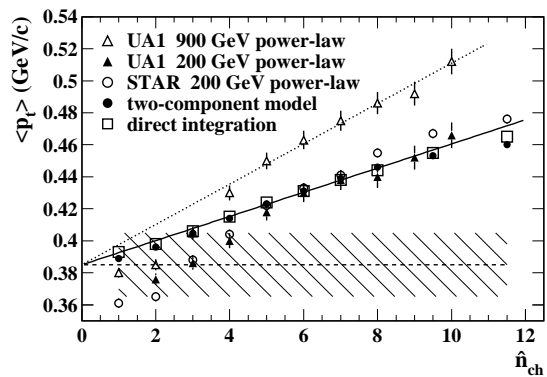}
\caption{\label{fig5}
$\langle p_t \rangle(\hat n_{ch})$ derived from the two-component $H_0$ gaussian amplitudes (solid dots), from the running integrals (open triangles) and from power-law fits to STAR and UA1 data (open circles, triangles). The solid and dotted lines correspond to Eq.~(\ref{mnpt}) with $\alpha =$ 0.0095 and 0.015 respectively.
}
\end{figure}

In Fig.~\ref{fig5} $\langle p_t \rangle(\hat n_{ch})$ values inferred from power-law fits to corrected STAR spectra are represented by open circles, consistent with a 200 GeV UA1 power-law analysis plotted as solid triangles~\cite{ua1}, but inconsistent at smaller $\hat n_{ch}$ with the two-component result from this analysis plotted as solid points. The two-component data were obtained with the $n_h/n_s$ values plotted as solid dots in Fig.~\ref{fig4a} (right panel). The solid line represents the two-component {analytic expression} for $\langle p_t \rangle$ in Eq.~(\ref{mnpt}) with $\alpha = 0.0095$. The UA1 results for 900 GeV~\cite{ua1} are plotted as open triangles. The dotted line corresponds to Eq.~(\ref{mnpt}) with $\alpha = 0.015$. $\langle p_t \rangle$ values obtained by {\em direct integration} of the extrapolated spectra are represented by the open squares.  The hatched region represents the {common} uncertainty in all means due to uncertainty in the particle yield in $p_t < 0.2$ GeV/c.


The $\langle p_t \rangle(\hat n_{ch})$ values in Fig.~\ref{fig5} obtained by direct integration of extrapolated spectra provide the best estimate of the physical trend. The results at 200 GeV for direct integration, the two-component model and power-law fits are consistent within errors for $\hat n_{ch} > 4$. The notable deviation of the power-law results from the two-component linear trend for $\hat n_{ch} < 5$ can be explained by the third-component structures at small $y_t$ and small $\hat n_{ch}$ in Fig.~\ref{fig3} (left panel). Those structures strongly influence (bias) {extrapolation} of the power-law function into the unmeasured region in $p_t < 0.2$ GeV/c so as to overestimate the inferred yield there (nominally 30\% of the total spectrum). The overestimate at small $p_t$ produces a sharp reduction of $\langle p_t \rangle(\hat n_{ch})$ values inferred from power-law fits. The additional yield at small $y_t$ in Fig.~\ref{fig3} itself corresponds to $\langle p_t \rangle \sim$ 0.4 GeV/c, and thus cannot physically lower the composite $\langle p_t \rangle$ below $\langle p_t \rangle_{soft} =$ 0.385 GeV/c. These $\langle p_t \rangle$ results demonstrate that the UA1 data {\em are} sensitive to the small-$y_t$ and small-$\hat n_{ch}$ structures revealed in this analysis when the more integral spectrum measure $\langle p_t \rangle$ is used. 


\section{Errors}



The statistical errors for the basic $y_t$ spectra in Fig.~\ref{fig1} are best indicated by the error bars on the difference distributions of Fig.~\ref{fig3} (left panel). That figure also compares the point-wise statistical errors to the hard-component structure inferred in this study, which is statistically well determined for all $n_{ch}$ classes. Monte Carlo calculations of background corrections with full detector response simulation are computer intensive. Because of limited statistics the statistical fluctuations in the Monte Carlo data used for background corrections are injected into the corrected data spectra as visible systematic errors: long-wavelength systematic error is reduced at the expense of increased short-wavelength random `systematic' error. Those errors are apparent as the nonstatistical short-wavelength structures in Figs.~\ref{fig1a} and~\ref{fig3a}.  The systematic uncertainties in the corrected spectra can be divided into $n_{ch}$-dependent and $n_{ch}$-independent uncertainties.

$n_{ch}$-independent systematic uncertainties include uncertainties in the corrections for tracking efficiency, backgrounds (mainly weak decays) and momentum resolution. Systematic spectrum corrections for this analysis were 20\% or less, except for the lowest two $p_t$ bins where they increased to 40\%. Statistical errors for the systematic corrections were typically less than 1\% (except as noted above for the background corrections). We estimate the uncertainties in the systematic corrections as 10\% of the correction values. The total uncertainty for the systematic corrections is then less than 2\% above $p_t =$ 0.4 GeV/c. The UA1 corrected $n_{ch}$-inclusive $p_t$ spectrum for 200 GeV \=p-p collisions~\cite{ua1}
agrees with the corresponding inclusive spectrum from the present analysis at the 2\% level. 

$n_{ch}$-dependent systematic errors could result from $n_{ch}$-dependent tracking inefficiencies. However, track detection and $p_t$ measurement in this analysis required no reference to other tracks or a fitted event vertex, thus minimizing any $n_{ch}$ dependencies. In effect, each track was treated in isolation independent of its relationship to any event, except for the timing requirement with the CTB. The tracking efficiencies for low-multiplicity (1-4) and high-multiplicity ($> 4$) events integrated over the $p_t$ acceptance were found to be consistent to 3\%, with a 1\% statistical error. We take that as an estimate of the $n_{ch}$-dependent systematic uncertainty.


The main source of systematic uncertainty in the shape of the hard-component structures isolated in Fig.~\ref{fig3} is the definition of $S_0$ as the lowest element of the regular sequence in Fig.~\ref{fig2} (left panel). $S_0$ is a rapidly-decreasing function in the interval $y_t =$ 1.6-3. The main effect of varying either $\beta_0$ or $n$ in $S_0$ is to change the magnitude of $S_0$ in that interval, shape changes being secondary. It is consistent within the two-component context to require that 1) component $H(n_{ch},y_t)$ be non-negative, placing an upper bound on $S_0$ in Fig.~\ref{fig3} and 2) that any $n_{ch}$-{\em independent} aspect of the distributions in Fig.~\ref{fig3} be minimized, determining a lower bound. Those criteria place stringent constraints on $S_0$ already in $y_t \sim$ 1.6-2, limiting systematic offsets at $y_t = 2$ to $\pm0.002$, the allowed range rapidly decreasing above that point according to the $S_0$ curve in Fig.~\ref{fig4} (right panel). The systematic uncertainty estimate corresponding to those trends is represented by the hatched region in Fig.~\ref{fig3} (right panel).


The nonstatistical power-law fit residuals in Fig.~\ref{fig1a} are as much as thirty times the statistical error. One of the findings of this study is that the power-law model function is inappropriate for these $p_t$ spectra. Systematic uncertainties for the fit parameters are therefore not meaningful. 

The fitting uncertainties for the fixed-model parameters are given at the bottom of Table~\ref{tab1}. Those uncertainties are meaningful relative to the fitting procedure defined in the two-component model context. The ability of that model to describe the data is apparent in Fig.~\ref{fig3a} (left panel). The only significant residuals correspond to a low-$y_t$ spectrum element (for $\hat n_{ch} = $ 1-4) deliberately omitted from the two-component model. One source of systematic uncertainty in those parameters is whether the fixed-model prescription forces a certain result by excluding some other which may better describe the data.

To test that possibility a free $\chi^2$ fit with all model parameters varying was conducted. The difference in the two cases is summarized in Table~\ref{tab1} and Fig.~\ref{fig4a} (left panel). In particular, there are substantial differences in the L\'evy exponent and the hard-component amplitude for the free fit depending on whether the position of the hard-component gaussian is constrained or not. When the gaussian position is constrained the free fit and the fixed model agree within the systematic uncertainties in the latter. The differences in the unconstrained fit are traced to significant departures of the shape of the hard component data peak from the symmetric gaussian peak shape: the model function could be further refined by adding a skewness (expected for fragmentation functions) to improve the stability of the fits. However, it is not our purpose to develop a complex representation of $y_t$ spectra, but rather to demonstrate the essential two-component aspects of the spectra with the simplest possible model function. The differences in fit parameters in Fig.~\ref{fig4a} (left panel) can therefore be taken as a generous estimate of  the systematic uncertainty in the fixed-component parameterization.





\section{Identified Particles}

Model functions $S_0$ and $H_0$ derived from this analysis of unidentified particles represent physical spectrum components $S$ and $H$ for several hadron types, mainly $\pi$, K and p. Two questions emerge: 1) to what extent do  $S_0$ and $H_0$ correspond to individual hadron types, and 2) to what extent does the $n_{ch}$ dependence of the $p_t$ spectrum truly separate  two physical components  $S$ and $H$? The soft component of one hadron species may have significant $n_{ch}$ dependence which could be misinterpreted as the hard component of another species, or of the combination of unidentified hadrons in this study. 

We can obtain some answers to those questions from $n_{ch}$-inclusive spectrum studies of identified hadrons. $p_t$ spectra for $\sqrt{s_{NN}}=200$ GeV p-p collisions have been measured for identified pions, kaons and protons~\cite{tofr}. Because $\hat n_{ch} \sim 1$ and the $p_t$ acceptance was [0.3,3] GeV/c
for that analysis the measured multiplicity-inclusive $p_t$ spectra are reasonably described by L\'evy distribution $S_0$, especially the kaon and proton spectra. The common L\'evy exponent for the three species is $n = 16.8\pm0.05$, compared to $n = 12.8\pm0.15$ measured in this analysis for unidentified hadrons. The slope parameter for identified pions is $T = 0.145\pm.001$ GeV, whereas for both kaons and protons $T = 0.23\pm0.005$ GeV, compared to $T = 0.1445\pm0.001$ GeV for unidentified hadrons in this analysis. 

The trend of $S_0$ with hadron species is easily understood. Addition of the `hotter' K and p spectra to the `cooler' pion spectrum flattens the unidentified hadron composite at larger $p_t$, reducing the exponent of $S_0$ to $n = 12.8$. At smaller $p_t$ the pion fraction dominates the composite spectrum, and the unidentified-hadron slope parameter is the same as the pion slope parameter. The effect of the heavier hadrons on the composite spectrum is mainly to reduce the L\'evy exponent from the larger physical value common to all three hadron species.


Information on the $n_{ch}$ dependence of $p_t$ spectra for identified particles is limited. A preliminary analysis of $K_0^s$ and $\Lambda$ $p_t$ spectra up to 4 GeV/c~\cite{strange} suggests that the  $n_{ch}$ dependence of both spectra can be described by a modest (5\%) reduction of $n$ with increasing $n_{ch}$. That trend can be compared to the free $\chi^2$ fit results for $S_0$ in Table~\ref{tab1} as shown in Fig.~\ref{fig4a} (left panel): $n$ increases by about 25\% over the measured $\hat n_{ch}$ range. 
That increase is traced to an attempt by the model to accommodate a skewness of the hard component in the data, not a true variation in the soft component.  

\section{Pythia Monte Carlo}

\begin{figure}[h]
\includegraphics[width=3.3in,height=1.65in]{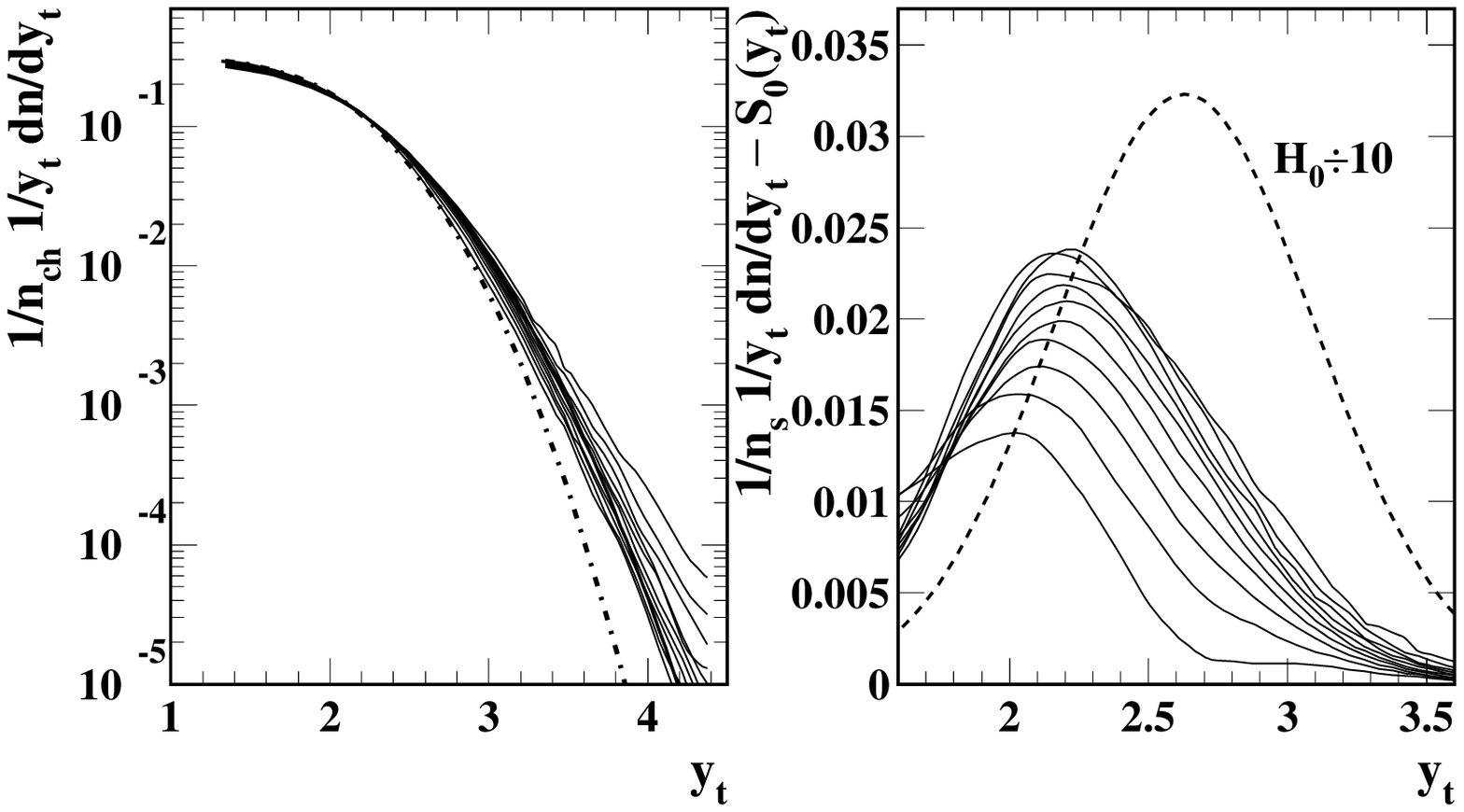}
\includegraphics[width=3.3in,height=1.65in]{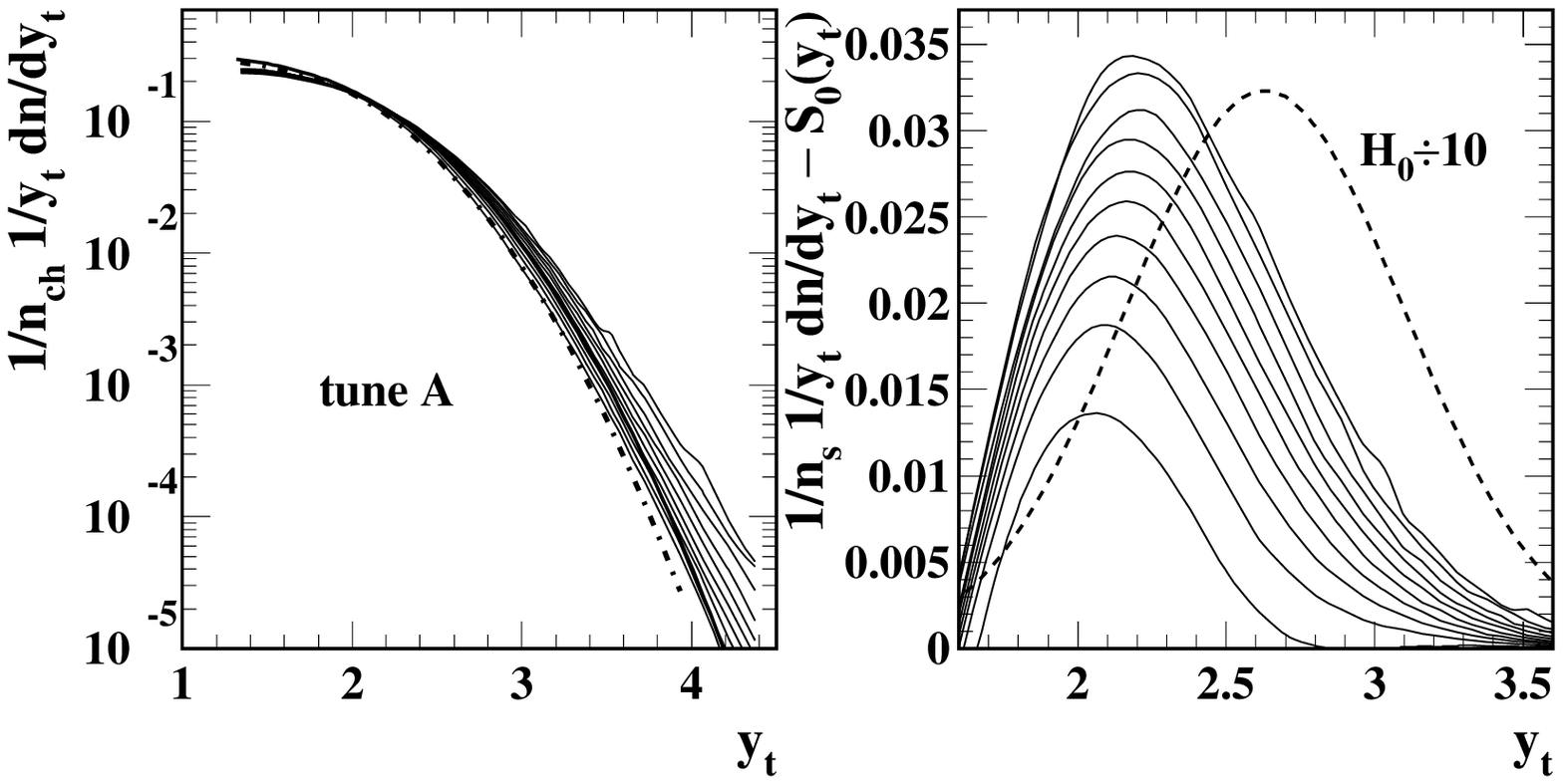}
\caption{\label{fig6a}
Two-component analysis applied to Pythia Monte Carlo data with the same multiplicity classes as for STAR data. The left and right panels may be compared with Figs. \ref{fig2a} (left panel) and \ref{fig3} (left panel) respectively. Dashed curves in the right panels represent the STAR data hard component for $\hat n_{ch} \sim 11$ ($H_0/10$).  Dash-dot curves in the left panels represent soft component $S_0$ optimized for each Monte Carlo configuration:  Pythia V6.222 default with parameters $T_0 = 0.147$ GeV and $n = 23$ (upper panel);  Pythia Tune A with parameters $T_0 = 0.137$ GeV and $n = 14$ (lower panel).
}
\end{figure}

A similar analysis of p-p collisions from the Pythia Monte Carlo~\cite{pythia} reveals substantial deviations from data. We studied default Pythia-V6.222 and Pythia 'tune A' (increased initial-state radiation and multiple soft parton scatters relative to the default) with parameters derived from studies of the underlying event in triggered jet events~\cite{under}. In Fig.~\ref{fig6a} (left panels) we show Pythia $p_t$ spectra normalized to unit integral and soft  reference $S_0$ (dash-dot curves) determined by the same criteria applied to STAR data. Those plots can be compared to Fig.~\ref{fig2a} (left panel). In Fig.~\ref{fig6a} (right panels) we show the results of subtracting reference $S_0$ from the normalized spectra in the left panel divided by $n_s / n_{ch}$. Those plots can be compared to Fig.~\ref{fig3} (left panel). The dashed curves are the hard component $H_0$ for STAR data divided by 10 to provide a reference. 

The $S_0$ parameters for Pythia-V6.222 in the upper panels are $T_0 = 0.147$ GeV and $n = 23$. The large value of $n$ implies that the Pythia soft component is nearly Maxwell-Boltzmann, in sharp contrast to RHIC data. The exponent is strictly limited to a large value by the Pythia data for $y_t < 2.5$. The $S_0$ parameters for Pythia tune A in the lower panels are $T_0 = 0.137$ GeV and $n = 14$. The smaller value of $n$ is comparable to the value 12.8 observed for RHIC data. 

The hard-component yield for Pythia is generally a factor of two to three less than the data (most apparent above $y_t  =2.7$), broader and peaked at a smaller value of $\bar y_t$. Pythia-V6.222 shows a saturating of the hard-component amplitude with increasing $\hat n_{ch}$, whereas Tune A shows a more uniform and significantly greater rate of increase. The large gaussian-shaped offset {\em common to all curves} and centered at $y_t \sim 2$ is also not observed in the data. That structure cannot be accommodated by the L\'evy distribution. The two Pythia Monte Carlos thus exhibit some features which agree qualitatively with experimental data but are quantitatively different. Tune A is closer to data than the default for soft and hard components, but the $n_{ch}$-independent gaussian offset near $y_t = 2$ persists and is not observed in the data.

\section{Discussion} \label{discussion}


A description of p-p collisions in terms of soft and hard components is natural at RHIC energies where significant hard parton scattering occurs but the underlying event~\cite{under} is still relatively simple. 
The two-component model of nuclear collisions can be applied to 1) the event-frequency distribution on $n_{ch}$ (two or more negative-binomial distributions)~\cite{tcomp1,duke}, 2) the dependence of $\langle p_t \rangle$ on $n_{ch}$~\cite{ua1,e735,knard}, 3) triggered jet correlations on $(\eta,\phi)$ (correlations from soft and hard event classes)~\cite{cdf} and 4) the $n_{ch}$ dependence of the $p_t$ or $y_t$ spectrum shape~\cite{e735}. The common theme is the relation of hard parton scattering to event multiplicity in the context of a `soft' underlying event. This paper emphasizes analysis type 4) -- study of the $n_{ch}$ dependence of the spectrum shape on transverse momentum $p_t$ and transverse rapidity $y_t$. 


Model functions $S_0(y_t)$ and $H_0(y_t)$ in Eq.~(\ref{2compalg}) can be viewed as the lowest-order elements of a perturbative expansion of the spectrum shape. Multiplicities $n_s(\hat n_{ch})$ and $n_h(\hat n_{ch})$ can  be interpreted as estimating the mean numbers of soft- and hard-component particles per event for a given $\hat n_{ch}$. The claim of simplicity for the two-component fixed model is supported by the small number of parameters, the simplicity of the model functions, the demonstration of necessity in Sec.~\ref{chi2} and the demonstration with residuals plots that there is no additional information in the spectra (aside from the small-$y_t$ `third component' which may represent additional physics).

We cannot rule out additional components or changes in the shapes of {\em physical} components $S$ and $H$. Each should be $n_{ch}$-dependent at some level, but the present analysis indicates that within the observed $\hat n_{ch}$ interval any such dependence is near the level of statistical error. A change in $S$ is suggested by the $n$ dependence of the free $\chi^2$ fits in Fig.~\ref{fig3a} (left panel). However, that behavior may simply be due to a coupling of soft and hard amplitudes in the free fit, with no physical significance.

A significant change in $H$ {\em is} expected at larger $\hat n_{ch}$ based on known jet physics: larger fragment multiplicities are produced by more energetic partons, with fragment distributions shifted to larger $y_t$~\cite{fragfunc}. Thus, the mean and width of $H$ should increase with $\hat n_{ch}$ at some point, but such changes are not observed beyond statistics within the $y_t$ and $\hat n_{ch}$ acceptances of this study. Apparently, the multiplicity increase in this analysis is dominated by increased {\em frequency of events with a single hard scattering} within a multiplicity class rather than bias toward more energetic partons. That scenario is consistent with the two-component model of~\cite{tcomp1}.


The soft-component L\'evy distribution $S_0 \equiv A_s/(1+\beta_0\, (m_t - m_0)/n)^n$~\cite{wilke} is similar in form to power-law function $A / (1 + p_t / p_0)^n$. However, the physical interpretations are quite different. The L\'evy distribution describes a nominally exponential function with a control parameter ({\em e.g.,} slope parameter) which undergoes gaussian-random fluctuations. Inverse exponent $1/n$ then measures the {\em relative variance} $\sigma^2_\beta / \beta_0^2$ of the control parameter~\cite{mtxmt}. In the limit $1/n \rightarrow 0$ the L\'evy distribution on $m_t$ becomes a true Maxwell-Boltzmann distribution. Those properties suggested the L\'evy distribution as a reference function for this analysis. Ironically, the `power-law' function in the form of a L\'evy distribution describes the soft component, not hard parton scattering. The L\'evy parameters can be interpreted in the context of an ensemble of hadron emitters with random transverse speeds, thermal radiation from moving sources as described by the Cooper-Frye formalism~\cite{cooper}. The expected QCD hard-scattering power-law trend is not evident in the data out to $p_t \sim 6$ GeV/c. 

In Fig.~\ref{fig1a} (right panel) we plot exponent $n$ values from power-law fits to data (solid points) and to the two-component fixed model (open circles) for the full range of $\hat n_{ch}$. The latter procedure simulates a power-law fit to data with no small-$y_t$ excursions or `third component' and illustrates the effect of those features on the exponent. The range of variation of the power-law exponent, in contrast to the two-component fixed model, and the substantial effect of the `third component' further illustrate that the power-law parameterization is sensitive to aspects of spectra inconsistent with its theoretical motivation, making fit results difficult to interpret physically. 



The gaussian shape of $H_0(y_t)$ inferred from this analysis can be compared with {\em fragmentation functions} from jet analysis of p-p, e-p and e-e collisions plotted on logarithmic variable $\xi_p \equiv \ln\{p_{jet}/p_{fragment}\}$, which also have an approximately gaussian shape~\cite{opal} explained in a QCD context as the interplay of parton splitting or branching at larger $p_t$ and the nonperturbative cutoff of the branching process at smaller $p_t$ due to gluon coherence~\cite{dok,fong}. The gaussian parameters are predicted by the pQCD MLLA (modified leading-log approximation)~\cite{mlla}. The hard component obtained in this analysis then represents not fragmentation functions from reconstructed large-$E_t$ jets but rather the average of a {\em minimum-bias} ensemble of fragmentation functions dominated by {\em low}-$Q^2$ parton scatters ($Q < 10$ GeV). In that context $H_0$ represents  {\em minimum-bias} partons dominated by {\em minijets}~\cite{minijet}. A previous study of small-$E_t$ clusters in 200 GeV p-p collisions~\cite{ua1a} suggested that semi-hard parton scattering or gluon radiation from projectile constituent quarks could produce substantial small-$p_t$ structure in hadron spectra similar to the hard component of this study.

{
A recent analysis of $p_t$ spectra in the interval 0.3 -- 10 GeV/c
for identified particles in p-p and d-Au collisions~\cite{pid2} used
the relativistic-rise particle identification scheme to extend the
spectra with very good  statistics to large $p_t$. That paper
compared the spectra to several NLO pQCD calculations and compared the
$m_t$ spectra of pions and protons. It concluded that there is
a transition region from soft to hard particle-production processes
at $p_t \sim$ 2 GeV/c in inclusive particle production, which would
appear to contradict the present results. However, the identified-particle spectra in that study below $p_t = 2.5$ GeV/c are
from a previous study~\cite{tofr} in which the point-to-point
systematic errors and the statistical errors are quite large, the latter due to the small acceptance of the prototype ToF detector. The ToF-based studies of multiplicity-averaged p-p collisions are therefore not sensitive to the hard-component structure reported in this paper, the great majority of which falls below 2.5 GeV/c. The present study takes a new approach by comparing
large-statistics inclusive-hadron spectra in several multiplicity
bins. Since the hard component is relatively enhanced in high-multiplicity
events we are able to extend our investigation of the hard
component to low $p_t$ by studying the trend of that enhancement.
}

The relative frequency of hard scatters in p-p collisions is described by the fifth model parameter $\alpha \sim 0.01$, representing the nearly-linear dependence of $n_h/n_s$ on $\hat n_{ch}$. We relate the hard-component amplitude to the frequency of hard collisions ($f \equiv$ number of hard collisions per NSD p-p collision) as $n_h(\hat n_{ch}) = \alpha\, \hat n_{ch}\, n_{s}(\hat n_{ch}) =  f(\hat n_{ch})\cdot \bar n_{mj}$, with mean true event multiplicity $  \bar n_{ch} = 2.5$ in one unit of pseudorapidity and mean minijet multiplicity $\bar n_{mj} = 2.5\pm1$~\cite{jetmult}. We then estimate the observed frequency of hard scatters in $\sqrt{s} = 200$ GeV p-p collisions as $f = \bar n_h / \bar n_{mj} = 0.012 \pm 0.004$ observed hard scatters per NSD p-p collision per unit of pseudorapidity. In that interpretation multiplicity $\hat n_{ch}$ serves as a `trigger' for hard parton scattering, determining the fraction of hard-scattering events in a given multiplicity class and thus the relative amplitude of the hard spectrum component.  


\begin{figure}[h]
\includegraphics[keepaspectratio,width=3.3in]{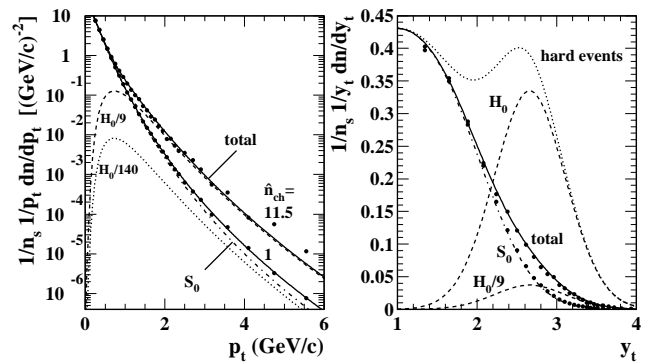}
\caption{\label{fig4}
Decomposition of inclusive $p_t$ spectra into a soft component represented by a L\'evy distribution on $m_t$ and a hard component represented by a gaussian on $y_t$. Dashed curves $H_0 / 9$ correspond to data for $\hat n_{ch} = 11.5$, while dotted curve $H_0 / 140$ correspond to data for $\hat n_{ch} = 1$ ({\em cf.} Fig.~\ref{fig1}). The dash-dot curves are soft reference $S_0$, and the solid curves are the totals of soft and hard components for the model. The dotted curve in the right panel estimates the shape of the inclusive $y_t$ distribution for those p-p collisions containing at least one minimum-bias hard parton scatter (hard events).
}
\end{figure}

Model functions $S_0$ and $H_0$ on $p_t$ and $y_t$ are summarized in Fig.~\ref{fig4}, which can be compared with Figs.~\ref{fig1}, ~\ref{fig2a} and ~\ref{fig3}. $H_0 / 9 \Rightarrow n_h / n_s = 0.11$ is compared to data for $\hat n_{ch} = 11.5$ and illustrates the role of the hard component in the measured spectra with sufficient amplitude to be visible in a linear plotting format (right panel). Similarly,  $H_0 / 140 \Rightarrow n_h / n_s = 0.007$ is compared to data for $\hat n_{ch} = 1$. Those coefficients are consistent with the measured hard-component gaussian amplitudes for $\hat n_{ch} = 1$ and 11.5 ({\em cf.} Fig.~\ref{fig4a} -- right panel). 

Collisions in the event ensemble containing at least one semi-hard parton scatter within the detector acceptance should have similar yields of soft and hard components (assuming an average minijet multiplicity of 2.5). The average $y_t$ spectrum for such hard events is illustrated by $S_0 + H_0$, shown as the dotted curve in  Fig.~\ref{fig4} (right panel). We cannot isolate such hard events in an unbiased manner, but we can infer their structure by extrapolating the $n_{ch}$ trends determined in this analysis. 

The left panel of Fig.~\ref{fig4} indicates the loss of visual sensitivity to spectrum structure when spectra are plotted on $p_t$. The hard component can appear to be a continuation of the soft component, whereas in the right panel the two components are clearly separate functional forms. $y_t$ provides a more balanced presentation of structure resulting from hard-scattered parton fragmentation, yet does not compromise study of the soft component, which is well-described by a simple error function on $y_t$~\cite{levy-err}. The transverse and longitudinal fragmentation systems undergo similar physical processes and should therefore be compared in equivalent plotting frameworks. Just as $y_z$ is preferred to $p_z$ we prefer $y_t$ to $p_t$.

\section{Summary}

In conclusion, we have studied the event multiplicity $n_{ch}$ dependence of high-statistics transverse momentum $p_t$ or transverse rapidity $y_t$ spectra from p-p collisions at $\sqrt{s} = 200$ GeV. We have determined that the `power-law' model function fails to describe the spectra for any $n_{ch}$, exhibiting large nonstatistical deviations from data. 
{ An earlier UA1 study reporting satisfactory power-law fits to data seems contradictory. However, it is statistically consistent with the present study because the UA1 data were derived from a much smaller event sample. 
}
We have analyzed the shapes of the spectra with a running-integral technique and determined that the spectra can be described precisely by a simple five-parameter model function. The algebraic model can in turn be related to a two-component physical model of nuclear collisions. 

The power-law function motivated by pQCD expectations for hard parton scattering better describes the soft component in the form of a L\'evy distribution on $m_t$ (two parameters). We observe for the first time that the hard component is well described by a {gaussian distribution on transverse rapidity $y_t$}, with shape approximately independent of multiplicity (two parameters). The hard-component multiplicity {\em fraction} increases almost linearly with event multiplicity (the fifth model parameter). 
{ 
A detailed comparison of (data $-$ model) residuals from the two-component fixed model and from free fits with all two-component model parameters varied confirms that the two-component fixed model is required by the data. 
}

The hard component may represent fragments from transversely scattered partons. The shape is consistent with fragmentation functions observed in LEP and PETRA $e^+$-$e^-$ and FNAL p-\=p collisions. The stability of the hard-component shape with event multiplicity suggests that a gaussian distribution on $y_t$ is a good representation of {minimum-bias} parton fragments. 
{ 
The relative abundance of soft and hard components at any $y_t$ of course depends on $y_t$ and $n_{ch}$, but most of the hard-component yield falls below 2.5 GeV/c.} There is evidence for a small but significant third component at smaller $y_t$ and smaller $n_{ch}$. Comparison with the Pythia Monte Carlo reveals qualitative differences from data.




We thank the RHIC Operations Group and RCF at BNL, and the
NERSC Center at LBNL for their support. This work was supported
in part by the Offices of NP and HEP within the U.S. DOE Office 
of Science; the U.S. NSF; the BMBF of Germany; IN2P3, RA, RPL, and
EMN of France; EPSRC of the United Kingdom; FAPESP of Brazil;
the Russian Ministry of Science and Technology; the Ministry of
Education and the NNSFC of China; IRP and GA of the Czech Republic,
FOM of the Netherlands, DAE, DST, and CSIR of the Government
of India; Swiss NSF; the Polish State Committee for Scientific 
Research; SRDA of Slovakia, and the Korea Sci. \& Eng. Foundation.

\begin{appendix}

\section{Symbol Definitions}

Below is a list of symbols and their definitions as used in  this paper.

\leftmargin .5in
\rightmargin .5in

\begin{quote}

\begin{description}
\itemsep .0in
\itemindent -.4in

\item[$y_t$:] transverse rapidity, replaces transverse momentum $p_t$ to provide improved visual access to fragment distributions
\item[$\hat n_{ch}$:] observed event multiplicity in the detector acceptance, also the event-class index
\item[$n'_{ch}$:] efficiency- and acceptance-corrected multiplicity in the detector acceptance
\item[$n_{ch}$:]  corrected and $p_t$-extrapolated or `true' multiplicity in the detector angular acceptance
\item[$n_s$:]  soft-component multiplicity in the acceptance
\item[$\tilde n_s$:] particular function of $\hat n_{ch}$ used to estimate $n_s$
\item[$n_h$:]  hard-component multiplicity in the acceptance: $n_s + n_h = n_{ch}$
\item[$\alpha$:] hard-component coefficient: $n_h / n_s \sim \alpha\, \hat n_{ch}$
\item[$S_0$:] unit-normal functional form of the soft component (L\'evy distribution on $m_t$)
\item[$N_0$:] running integral of soft reference $S_0$
\item[$H_0$:] unit-normal functional form of the hard component (gaussian distribution on $y_t$)
\item[$A,\,p_0,\,n$:] power-law model parameters
\item[$A_s,\,\beta_0,\,n$:] soft-component L\'evy distribution parameters, $1/\beta_0 = T_0$, the slope parameter
\item[$A_h,\,\bar y_t,\,\sigma_{y_t}$:] hard-component gaussian parameters


\end{description}
\end{quote}

\end{appendix}


\begin{thebibliography}{99}
\bibitem{yt} Transverse rapidity $y_t = \ln\{(m_t + p_t)/m_0\}$ in a longitudinally co-moving frame ($p_z = 0$) is equivalent in form to longitudinal rapidity $y_z = \ln\{(E+p_z)/m_0\}$ in a transversely co-moving frame ($p_t = 0$).
\bibitem{levy-err} The L\'evy distribution on $y_t$ is well approximated by an error function: $S_0(y_t;y_{t0},\xi_{y_t}) = A(y_{t0},\xi_{y_t})\, \{1+\text{erf}[-(y_t - y_{t0})/\xi_{y_t}]\}/2$. For these data  $A = 0.454\pm0.005$, $ y_{t0} = 2.03\pm 0.02$ and $\xi_{y_t} = 0.76 \pm 0.03$.  
\end{thebibliography}
\end{document}